\documentclass[aps,prd,preprintnumbers,groupedaddress,nofootinbib,amssymb,notitlepage,eqsecnum]{revtex4-2}
\usepackage{here}
\usepackage[dvipdfmx]{graphicx}
\usepackage{amsmath,amsthm,amssymb}
\usepackage{bm}
\usepackage{color}

\usepackage{amsfonts}
\usepackage{dcolumn}
\usepackage{hyperref}
\allowdisplaybreaks[1]
\usepackage{stackengine}

\newcommand{\be}{\begin{equation}}  
\newcommand{\ee}{\end{equation}}
\newcommand{\ba}{\begin{eqnarray}}
\newcommand{\ea}{\end{eqnarray}}

\newcommand{\rd}{{\rm d}}

\newcommand{\bem}{\begin{bmatrix}}
\newcommand{\eem}{\end{bmatrix}}
\newcommand{\Mpl}{M_{\rm Pl}}

\allowdisplaybreaks

\begin{document}

\preprint{YITP-24-109, WUCG-24-08}

\title{Scrutinizing black hole stability in cubic vector Galileon theories}

\author{Antonio De Felice$^{a}$}
\email{antonio.defelice@yukawa.kyoto-u.ac.jp}  

\author{Ryotaro Kase$^{b}$}
\email{r.kase@rs.tus.ac.jp}  

\author{Shinji Tsujikawa$^{c}$}
\email{tsujikawa@waseda.jp} 

\affiliation{$^{a}$Center for Gravitational Physics and Quantum Information, Yukawa Institute for Theoretical Physics, Kyoto University, 606-8502, Kyoto, Japan\\
$^{b}$Department of Physics, Faculty of Science, 
Tokyo University of Science, 1-3, Kagurazaka,
Shinjuku-ku, Tokyo 162-8601, Japan\\
$^{c}$Department of Physics, Waseda University, 3-4-1 Okubo, 
Shinjuku, Tokyo 169-8555, Japan}

\begin{abstract}

In a subclass of generalized Proca theories where a cubic vector Galileon term breaks the $U(1)$ gauge invariance, it is known that there are static and spherically symmetric black hole (BH) solutions endowed with nonvanishing temporal and longitudinal vector components. Such hairy BHs are present for a vanishing vector-field mass ($m=0$) with a non-zero cubic Galileon coupling $\beta_3$. 
We study the linear stability of those hairy BHs by considering even-parity perturbations in the eikonal limit. 
In the angular direction, we show that one of the three dynamical perturbations has a nontrivial squared propagation speed $c_{\Omega,1}^2$, while the other two dynamical modes are luminal. 
We could detect two different unstable behaviors of perturbations in all the parameter spaces of hairy asymptotically flat BH solutions we searched for. 
In the first case, an angular Laplacian instability on the horizon is induced by negative $c_{\Omega,1}^2$.
For the second case, it is possible to avoid this horizon instability, but in such cases, the positivity of $c_{\Omega,1}^2$ is violated at large distances. 
Hence these hairy BHs are generally prone to Laplacian instabilities along the angular direction in some regions outside the horizon. 
Moreover, we also encounter a pathological behavior of the radial propagation speeds $c_r$ possessing two different values of $c_r^2$ for one of the dynamical perturbations.
Introducing the vector-field mass $m$ to cubic vector Galileons, however, we show that the resulting no-hair Schwarzschild BH solution satisfies all the linear stability conditions in the small-scale limit, with luminal propagation speeds of three dynamical even-parity perturbations.
\end{abstract}

\date{\today}


\maketitle

\section{Introduction}
\label{introsec}

In General Relativity (GR), black hole (BH) solutions on a static and spherically symmetric (SSS) background are characterized by only two independent classical parameters, mass and charge \cite{Israel:1967wq, Carter:1971zc, Hawking:1971vc}. 
In the presence of extra degrees of freedom (DOFs) like a scalar or vector field, it is possible for  
BHs to have additional ``hairs'' in the form of a scalar or vector charge. 
For a scalar field $\phi$ that depends on a radial distance $r$ alone, the existence of hairy BHs 
is quite limited even in the context of scalar-tensor theories \cite{Hawking:1972qk, Bekenstein:1995un, Sotiriou:2011dz, Hui:2012qt, Graham:2014mda, Faraoni:2017ock}. 
Indeed, for most general scalar-tensor theories with second-order field equations of motion 
(Horndeski theories \cite{Horndeski:1974wa}), the realization of linearly stable and asymptotically flat hairy BH solutions requires that the scalar field is at least coupled to a Gauss-Bonnet term \cite{Creminelli:2020lxn, Minamitsuji:2022vbi} (see also Refs.~\cite{Kanti:1995vq, Torii:1996yi, Kanti:1997br, Sotiriou:2013qea, Minamitsuji:2022mlv}).
Allowing time dependence for the scalar-field profile while keeping the time-independent metric, there are possibilities of realizing stealth-type BH solutions with nontrivial scalar profiles \cite{Arkani-Hamed:2003pdi, Mukohyama:2005rw, Babichev:2013cya, Kobayashi:2014eva, Motohashi:2019ymr, DeFelice:2022xvq}.

If we consider a vector field $A_{\mu}$ with the field strength $F_{\mu \nu}=\partial_{\mu}A_{\nu}-\partial_{\nu}A_{\mu}$, Einstein-Maxwell theory with the electromagnetic Lagrangian $F=-(1/4)F_{\mu \nu}F^{\mu \nu}$ gives rise to a Reissner-Nordstr\"om (RN) BH whose metric contains the electric or magnetic charge besides the BH mass.
Einstein-Maxwell theory respects the $U(1)$ gauge invariance under a shift $A_{\mu} \to A_{\mu}+\partial_{\mu} \chi$. 
The $U(1)$ gauge symmetry is broken by introducing a mass $m$ of the vector field with the Lagrangian $m^2 X$, where $X=-(1/2)A_{\mu}A^{\mu}$.
This Proca theory leads to the propagation of a longitudinal scalar mode besides two transverse 
vector polarizations. 
In Einstein-Proca theory where the gravity sector is described by GR with the massive Proca field, the BH solutions on the SSS background reduce to the Schwarzschild geometry with vanishing temporal and longitudinal vector components \cite{Bekenstein:1971hc, Bekenstein:1972ky}. 

Derivative self-interactions of the vector field containing the $X$ dependence also break the 
$U(1)$ gauge symmetry in general. 
To avoid the propagation of extra DOFs associated with Ostrogradsky ghosts \cite{Ostrogradsky:1850fid}, it is desirable to construct a generalized version of Proca theories with second-order field equations of motion. 
Analogous to scalar Horndeski theories, one can also implement couplings between the vector field and gravity. 
The authors of Refs.~\cite{Heisenberg:2014rta, Tasinato:2014eka, Allys:2015sht, BeltranJimenez:2016rff, Allys:2016jaq} constructed the Lagrangian of such generalized Proca (GP) theories, in which there are five propagating DOFs in total (one longitudinal scalar, two transverse vectors, and two tensor polarizations).
Einstein-Proca theory is a specific case of GP theories given by the Lagrangian ${\cal L}=F+m^2 X+\Mpl^2 R/2$, where $\Mpl$ is the reduced Planck mass and $R$ is the Ricci scalar. 

If we apply GP theories to BH solutions on the SSS background, it is possible to realize hairy BHs in the presence of cubic Lagrangians of the form $G_3(X) \nabla_{\mu}A^{\mu}$ \cite{Heisenberg:2017xda, Heisenberg:2017hwb}, where $G_3$ is a function of $X$ and $\nabla_{\mu}$ is a covariant derivative operator. 
The linear coupling $G_3(X)=\beta_3 X$, where $\beta_3$ is a constant, is in analogy with the scalar Galileon by taking the limit $A_{\mu} \to \nabla_{\mu} \phi$, under which the Lagrangian $\beta_3 X \nabla_{\mu}A^{\mu}$ reduces to $\beta_3 X_\phi \square \phi$ with $X_\phi\equiv-(1/2)\nabla_{\mu} \phi \nabla^{\mu} \phi$ and $\square=\nabla_{\mu}\nabla^{\mu}$.
On the Minkowski background, the field equation of motion for scalar Galileons is invariant under the 
shift $\nabla_{\mu} \phi \to  \nabla_{\mu} \phi+b_{\mu}$, which is the reason for being called 
Galileons \cite{Nicolis:2008in,Deffayet:2009wt}. 
We call theories with the Lagrangian $\beta_3 X \nabla_{\mu}A^{\mu}$ cubic vector Galileons.

In a subclass of GP theories described by the Lagrangian ${\cal L}=F+\beta_3 X \nabla_\mu A^\mu+(\Mpl^2/2)R$, the derivative self-interaction of cubic vector Galileons gives rise to 
asymptotically flat hairy BHs with nonvanishing temporal and longitudinal components \cite{Heisenberg:2017xda, Heisenberg:2017hwb}.
Here, we would like to stress that the mass term $m^2 X$ is absent in the Lagrangian and hence the vector Galileon self-interaction $\beta_3 X \nabla_\mu A^\mu$ breaks the $U(1)$ gauge symmetry.
In such theories, the field equation of motion for the longitudinal vector component $A_1$ can be written in the form $\beta_3 {\cal F}(A_1, A_0, A_0',\cdots)=0$, where ${\cal F}$ 
contains $A_1$, the temporal component $A_0$ and its radial derivative $A_0'$, as well as metric components $f$, $h$ and their radial derivatives.
So long as $\beta_3 \neq 0$, there are nonvanishing solutions to $A_1$ which are expressed in terms of $A_0$, $f$, $h$, and their radial derivatives.
In such cases, a scalar charge appears in $A_1$ besides an electric charge in $A_0$. 
This gives rise to a background metric different from the RN or Schwarzschild type.
We note that the scalar Galileon for a radial-dependent profile $\phi(r)$ does not possess a corresponding hairy BH solution \cite{Hui:2012qt, Minamitsuji:2022vbi}.

To understand whether the hairy BH induced by cubic vector Galileons does not pose theoretical pathologies such as the appearance of ghosts and Laplacian instabilities, we need to consider perturbations on the SSS background for both odd- and even-parity sectors.
The current state-of-the-art of BH perturbations in GP theories is that the linear stability conditions of odd-parity modes in the small-scale limit are known for general cases in which the vector field is coupled to the Ricci scalar, Einstein tensor, and double dual Riemann tensor \cite{Kase:2018voo} (see Refs.~\cite{Chagoya:2016aar, Fan:2016jnz, Minamitsuji:2016ydr, Cisterna:2016nwq, Chagoya:2017fyl, Babichev:2017rti, Heisenberg:2017xda, Heisenberg:2017hwb, Garcia-Saenz:2021uyv, Aoki:2023jvt} for BH solutions in the presence of couplings between the vector field and curvatures). 
In general, the odd-parity sector contains two dynamical DOFs arising from the vector field and the gravity sector, respectively.
The analysis of odd-parity perturbations allows one to exclude some of BH solutions, e.g., stealth Schwarzschild BHs realized by a quartic coupling $G_4(X) \supset \beta_4 X$ \cite{Chagoya:2016aar}, due to the instability of the vector-field perturbation around the horizon. For cubic vector Galileon theories mentioned above, the linear stability conditions in the odd-parity sector are always satisfied, with the luminal propagation speeds in both radial and angular directions \cite{Kase:2018voo}. 

In this paper, we will consider the propagation of even-parity perturbations to study the linear stability of hairy BHs realized by cubic vector Galileons. 
The even-parity sector contains one longitudinal scalar, one transverse vector, and one tensor polarization. 
Unlike theories with the $U(1)$ gauge invariance \cite{Moncrief:1974ng, Zerilli:1974ai, Moncrief:1975sb, Nomura:2020tpc, Kase:2023kvq, Chen:2024hkm}, it is generally nontrivial to identify the three dynamical DOFs of even-parity perturbations due to the possible mixture of the longitudinal scalar mode with the other two DOFs. 
As we will see in this paper, we should choose an appropriate gauge condition for this purpose.
By using the WKB-type approximation further, we can derive the linear stability conditions of three dynamical DOFs in the small-scale limit.

For hairy BHs induced by cubic vector Galileons, we will show that one dynamical even-parity perturbation has nontrivial propagation speeds along the angular and radial directions. 
In contrast, the other two dynamical DOFs propagate with the speeds of light. 
In all the asymptotically flat hairy solutions we have found, we identify two different behaviors: 1) either there is Laplacian instability around the BH horizon associated with a negative squared sound speed $c_{\Omega,1}^2=-2$ along the angular direction; 2) or $c_{\Omega,1}^2$ generally approaches negative values at large distances. 
Moreover, the squared radial propagation speed for one of the dynamical perturbations has two different values of $c_{r,1}^2$. Due to the angular Laplacian instability and pathological behavior of the radial propagation, the hairy BHs induced by cubic vector Galileons are excluded as stable configurations on the SSS background.

If we take into account a vector-field mass $m$ in addition to the cubic vector Galileon, i.e., in theories with the Lagrangian ${\cal L}=F+m^2X+\beta_3 X \square \phi+(\Mpl^2/2)R$, we will see that 
the resulting background BH solution is described by the Schwarzschild geometry without vector hairs ($A_0=0=A_1$). In this case, we will also study the linear stability of BHs against even-parity perturbations and show that the no-ghost conditions are satisfied outside the horizon with the luminal angular and radial propagation speeds for three dynamical DOFs. 
Thus, introducing a mass term to the cubic vector Galileon overcomes the instability problem of hairy BHs mentioned above. 

This paper is organized as follows. 
In Sec.~\ref{scasec}, we discuss BH solutions on the SSS background in theories with the Lagrangian ${\cal L}=F+m^2X+\beta_3 X \square \phi+(\Mpl^2/2)R$. 
In particular, we revisit the hairy BH solution present for $m=0$ and see how introducing the nonvanishing mass $m$ results in the no-hair Schwarzschild BH.
In Sec.~\ref{gaugesec}, we formulate BH perturbations in the even-parity sector for cubic vector Galileon theories and discuss several possible gauge choices. 
In Sec.~\ref{sta1sec}, we derive linear stability conditions of three dynamical perturbations by choosing a proper gauge and show that the angular Laplacian instability is generally present for hairy BHs realized by cubic vector Galileons with $m=0$. In Sec.~\ref{sta2sec}, we show that introducing a vector-field mass leads to no-hair BH solutions that suffer from neither ghost nor Laplacian instabilities. Sec~\ref{consec} is devoted to conclusions.

\section{Black holes with cubic vector Galileons}
\label{scasec}

Let us consider a covector field $A_{\mu}$ with the field strength $F_{\mu \nu}=\partial_{\mu}A_{\nu}-\partial_{\nu}A_{\mu}$. 
We introduce the two scalars
\be
F=-\frac{1}{4}F_{\mu \nu} F^{\mu \nu}\,,\qquad 
X=-\frac{1}{2}A_{\mu}A^{\mu}\,,
\label{FX}
\ee
where $F$ respects the $U(1)$ gauge invariance under the shift $A_{\mu} \to A_{\mu}+\partial_{\mu} \chi$. 
If we introduce a mass $m$ of the vector field, the Lagrangian $m^2 X$ breaks the $U(1)$ gauge symmetry. 
The generalization of massive Proca theories to the curved background, with the second-order property of field equations of motion maintained, is known as GP theories \cite{Heisenberg:2014rta, Tasinato:2014eka, Allys:2015sht, BeltranJimenez:2016rff, Allys:2016jaq}.
In this paper, we focus our attention on a subclass of GP theories given by the action 
\begin{equation}
{\cal S}=\int {\rm d}^{4}x \sqrt{-g} 
\left( F+m^2 X+\beta_3 X \nabla_{\mu}A^{\mu}
+\frac{\Mpl^2}{2}R \right)\,,
\label{action}
\end{equation}
where $g$ is a determinant of the metric tensor $g_{\mu\nu}$, and $\beta_3 X \nabla_{\mu}A^{\mu}$ is the Lagrangian of cubic vector Galileons with a coupling constant $\beta_3$.  
The full GP theories contain gravitational couplings with the vector field of the forms $G_4(X) R$, $G_5(X) G_{\mu \nu} \nabla^{\mu}A^{\nu}$, and $G_{6}(X)L^{\mu \nu \alpha \beta} \nabla_{\mu}A_{\nu} 
\nabla_{\alpha} A_{\beta}$ together with counter terms eliminating derivatives higher than second order, where $R$ is the Ricci scalar, $G_{\mu \nu}$ is the Einstein tensor, and $L^{\mu \nu \alpha \beta}$ is the double dual Riemann tensor.  
Since we would like to understand the stability of hairy BH solutions induced by cubic vector Galileons, we will consider theories with Eq.~(\ref{action}), in which both $G_5$ and $G_6$ are vanishing with $G_4=\Mpl^2/2$. From a phenomenological point of view, this implies that the speed of tensor perturbations is equivalent to that of light. 
Hence the theories automatically evade the bound on the speed of gravity constrained from gravitational-wave measurements \cite{LIGOScientific:2017zic}.

The line element of the SSS background is given by 
\begin{equation}
\rd s^{2} =-f(r) \rd t^{2} +h^{-1}(r)\rd r^{2} + 
r^{2} (\rd \theta^{2}+\sin^{2}\theta\, \rd \varphi^{2})\,,
\label{metric}
\end{equation}
where $f$ and $h$ are functions of the radial distance $r$.
We consider the vector-field profile $A_{\mu}=\left[ A_0(r), A_1(r), 0, A_{\varphi} (\theta) \right]$,
where $A_0$ and $A_1$ are functions of $r$ and $A_{\varphi}$ is a function of $\theta$. 
Then, the two scalar products in Eq.~(\ref{FX}) reduce to
\begin{equation}
F=\frac{h}{2f}A_0'^2-\frac{1}{2r^4 \sin^2 \theta} \left( 
\frac{\rd A_{\varphi}} {\rd \theta} \right)^2\,,\qquad 
X=\frac{A_0^2}{2f}-\frac{h A_1^2}{2} 
-\frac{A_{\varphi}^2(\theta)}{2r^2 \sin^2 \theta}\,,
\end{equation}
where a prime represents the derivative with respect to $r$.
For compatibility with the SSS background, we require that both $F$ and $X$ are functions of $r$ alone.
If $\rd A_{\varphi}/\rd \theta$ in $F$ is proportional to $\sin \theta$, $F$ depends only on $r$. Then, we obtain $A_{\varphi}=-q_M \cos \theta$, where $q_M$ is a constant corresponding to a magnetic charge. 
For $q_M \neq 0$, however, the last term in $X$ has the $\theta$ dependence of the form $-q_M^2/(2r^2 \tan^2 \theta)$. 
This means that, unless $q_M=0$, the presence of $X$-dependent terms in the action (\ref{action}) is incompatible with the spherical symmetry. 
Hence we will set $q_M=0$, i.e., $A_{\varphi}(\theta)=0$ in the whole background analysis given below.
Namely, we will consider the vector-field configuration 
\begin{equation}
A_{\mu}=\left[ A_0(r), A_1(r), 0, 0 \right]\,,
\label{vecback}
\end{equation}
with which $F=h A_0'^2/(2f)$ and $X=A_0^2/(2f)-hA_1^2/2$. 
In theories with the broken $U(1)$ gauge symmetry, the longitudinal component $A_1(r)$ is generally present besides the temporal component $A_0(r)$. 

On the SSS background (\ref{metric}), the gravitational equations of motion for $f$ and $h$ are given by 
\begin{eqnarray}
h' &=& \frac{2\Mpl^2 f(h - 1)+ 4 \beta_3 h r A_0^2 A_1 
+ [2 \beta_3 h A_0 A_0'A_1  + A_0^2 (m^2+2 \beta_3 h A_1') 
+ h \{ A_0'^2 + A_1^2 f (m^2-2\beta_3 h A_1') \}]r^2}
{r [\beta_3r A_1 (fh A_1^2- A_0^2) - 2 \Mpl^2 f]}\,,\label{back1}\\
f' &=& \frac{f[2\Mpl^2f(1 - h)+ \{ (2 \beta_3 A_0 A_1 - A_0')hA_0'
+ m^2 A_0^2 \} r^2 + A_1^2 rfh (4 \beta_3 h A_1 + m^2r)]}
{hr[ \beta_3 r A_1 (A_0^2-fh A_1^2)+ 2\Mpl^2 f]}\,.\label{back2}
\end{eqnarray}
The vector-field components obey the following equations 
\begin{eqnarray}
& &
A_0''+\left( \frac{2}{r}-\frac{f'}{2f}+\frac{h'}{2h} \right) A_0'
-\left[ \frac{m^2}{h}+\beta_3 \left\{  A_1'+\left( 
\frac{2}{r}+\frac{f'}{2f}+\frac{h'}{2h} \right)A_1 \right\} \right]A_0=0\,,
\label{back3}\\
& &
2m^2 r f^2 A_1+\beta_3 \left[ f h (rf' + 4f)A_1^2 - r f'A_0^2 
+ 2rf A_0 A_0' \right]=0\,.\label{back4}
\end{eqnarray}
We solve Eq.~(\ref{back4}) for $A_1$, as 
\begin{equation}
A_1=\frac{-m^2 r f^2 \pm 
\sqrt{rf [ m^4 r f^3 + \beta_3^2 ( rf' + 4f) h 
A_0 (A_0 f' - 2A_0' f)]}}{\beta_3 fh (rf' + 4f)}\,.
\label{back4d}
\end{equation}
So long as $\beta_3 \neq 0$, the longitudinal component $A_1$ is nonvanishing. We can integrate Eqs.~(\ref{back1})-(\ref{back3}) with Eq.~(\ref{back4d}) to solve for $h$, $f$, and $A_0$ for given boundary conditions on the horizon.

When $m=0$ and $\beta_3=0$, the theory respects the $U(1)$ gauge invariance without the longitudinal propagation ($A_1=0$).
Integrating Eqs.~(\ref{back1})-(\ref{back3}) with asymptotically flat boundary conditions $f \to 1$ and $h \to 1$ at spatial infinity, we obtain the RN solution characterized by 
\begin{equation}
f=h=1-\frac{2M}{r}+\frac{q^2}{2\Mpl^2 r^2}\,,\qquad 
A_0=P+\frac{q}{r}\,,\qquad A_1=0\,,\qquad 
({\rm for}~~m=0~~{\rm and}~~\beta_3=0)\,,
\end{equation}
where $M$, $q$, and $P$ are constants. 
The mass and charge of BHs correspond to $M$ and $q$, respectively.

Einstein-Proca theory corresponds to the nonvanishing mass $m \neq 0$ with $\beta_3=0$.   
In this case, we have $A_1=0$ from Eq.~(\ref{back4}). 
Moreover, combining Eq.~(\ref{back1}) with Eq.~(\ref{back2}) gives 
\begin{equation}
\left( \frac{f}{h} \right)'=
\frac{m^2 r A_0^2}{\Mpl^2 h^2}\,,
\qquad \quad
({\rm for}\quad \beta_3 =0)\,.
\label{fhre}
\end{equation}
In the vicinity of an outer horizon located at $r=r_h$, the metric components can be expanded as 
\begin{equation}
f=\sum_{i=1} f_i \left( r/r_h-1 \right)^i\,,\qquad 
h=\sum_{i=1} h_i \left(r/r_h-1 \right)^i\,,
\end{equation}
so that the left-hand side of Eq.~(\ref{fhre}) is finite at $r=r_h$. 
Since the denominator on the right-hand side of Eq.~(\ref{fhre}) contains the vanishing term $h^2$ on the horizon, we require that $A_0(r=r_h)=0$.
At spatial infinity, the left-hand side of Eq.~(\ref{fhre}) is also finite, and hence the condition $A_0 (r \to \infty)=0$ needs to be satisfied. 
Since the solution to Eq.~(\ref{back3}) contains a growing mode proportional to $e^{mr}/r$ at large distances, the deviation of $A(r)$ from 0 at an intermediate distance results in the increase $A(r) \propto e^{mr}/r$ for $r \gtrsim 1/m$. 
This means that $A_0(r)$ should vanish at any distance $r$. 
Then, we obtain the Schwarzschild solution with the vanishing vector field, 
\begin{equation}
f=h=1-\frac{2M}{r}\,,\qquad 
A_0=0\,,\qquad A_1=0\,,\qquad 
({\rm for}~~m \neq 0~~{\rm and}~~\beta_3=0)\,,
\label{Schwarz}
\end{equation}
and hence there is no charge $q$ in this case. 
This no-hair property of BHs in Einstein-Proca theory is consistent with the argument of Bekenstein \cite{Bekenstein:1971hc, Bekenstein:1972ky}.

Besides the two cases discussed above, there are two other cases: (A) $m=0$ and $\beta_3 \neq 0$, 
and (B) $m \neq 0$ and $\beta_3 \neq 0$, which we will discuss in turn.

\subsection{$m=0$ and $\beta_3 \neq 0$}
\label{hairBHsec}

In this case, the existence of a hairy BH solution with $A_1 \neq 0$ was first recognized in Ref.~\cite{Heisenberg:2017xda}. 
As we observe in Eq.~(\ref{back4}), this hairy solution does not exist for $\beta_3=0$. 
In other words, we cannot recover the RN solution by taking the continuous limit $\beta_3 \to 0$.
The solutions expanded at spatial infinity and around the horizon were also derived in Ref.~\cite{Heisenberg:2017hwb} by imposing the boundary conditions $f \to 1$ and $h \to 1$ as $r \to \infty$.
While the property $h (r \to \infty)=1$ follows by the background equations of motion, the asymptotic value of $f$ is usually arbitrary due to the time reparametrization freedom in the metric (\ref{metric}). 
Here, we allow the possibility of having an asymptotic value $f_{\infty}=f(r \to \infty)$ different from 1. 
Indeed, we will show that the reparametrization of $f$ to 1 at infinity is possible by properly rescaling the temporal vector component.

At spatial infinity, we will perform the following expansions 
\begin{equation}
f=f_{\infty}+\sum_{i=1}^{\infty} \frac{\tilde{f}_i}{r^i}\,,\qquad 
h=h_{\infty}+\sum_{i=1}^{\infty} \frac{\tilde{h}_i}{r^i}\,,\qquad 
A_0=P+\sum_{i=1}^{\infty} \frac{\tilde{a}_i}{r^i}\,,\qquad 
A_1=\tilde{b}_0+\sum_{i=1}^{\infty} \frac{\tilde{b}_i}{r^i}\,,
\label{largeex}
\end{equation}
where $f_{\infty}$, $\tilde{f}_i$, $h_{\infty}$, $\tilde{h}_i$, $P$, $\tilde{a}_i$, $\tilde{b}_0$, and $\tilde{b}_i$ are constants. 
For consistency with the background Eqs.~(\ref{back1})-(\ref{back4}), we require that $h_{\infty}=1$, $\tilde{f}_1=f_{\infty}\tilde{h}_1$, $\tilde{f}_2=0$, $\tilde{b}_0=0$, and $\tilde{b}_1=0$. 
Setting $\tilde{h}_1=-2M$ and $\tilde{b}_2=q_s$, the solutions expanded up to the order $1/r^4$ in Eq.~(\ref{largeex}) are given by
\begin{eqnarray}
f &=& f_{\infty} \left[ 1-\frac{2M}{r}-\frac{P^2 M^3}{6\Mpl^2 f_{\infty}r^3}
+\frac{M^4 P^2 (P^2-2\Mpl^2 f_{\infty})+3\Mpl^2 f_{\infty}^2 q_s^2}
{3\Mpl^2 f_{\infty} (2\Mpl^2 f_{\infty}-P^2)r^4} 
+{\cal O}(r^{-5}) \right]\,,\label{flar}\\
h &=& 1-\frac{2M}{r}-\frac{P^2 M^2}{2\Mpl^2 f_{\infty}r^2}
-\frac{P^2 M^3}{2\Mpl^2 f_{\infty}r^3}
+\frac{2M^4 P^2 (P^2-2\Mpl^2 f_{\infty})+12\Mpl^2 
f_{\infty}^2 q_s^2}
{3\Mpl^2 f_{\infty} (2\Mpl^2 f_{\infty}-P^2)r^4} 
+{\cal O}(r^{-5})\,,\\
A_0&=& 
P-\frac{PM}{r}-\frac{PM^2}{2r^2}
-\frac{PM^3(P^2+6\Mpl^2 f_{\infty})}
{12\Mpl^2f_{\infty}r^3}
-\frac{P^2M^4(2P^2+5\Mpl^2f_{\infty})
(P^2-2\Mpl^2f_{\infty})+8\Mpl^4 f_{\infty}^3q_s^2}
{8P\Mpl^2f_{\infty}(P^2-2\Mpl^2f_{\infty})r^4} \nonumber \\
& &
+{\cal O}(r^{-5})\,,\\
A_1&=&
\frac{q_s}{r^2}+
\frac{M(M+2q_s \beta_3)}{\beta_3 r^3}
+\frac{12M^3\Mpl^2 f_{\infty}+q_s M^2 
(P^2+16\Mpl^2 f_{\infty})
\beta_3}{4\beta_3 \Mpl^2  f_{\infty}r^4}+{\cal O}(r^{-5})\,.
\label{fhA0}
\end{eqnarray}
Note that $M$, $-PM$, and $q_s$ correspond to the mass, electric charge, and scalar charge carried by the BH, respectively. For $f_{\infty}$ different from 1, we can perform the following rescalings
$f(r) \to \tilde{f}(r)\,f_\infty$, $P \to \tilde{P}\sqrt{f_\infty}$, and $A_0(r) \to \tilde{A}_0(r)\sqrt{f_\infty}$, so that $\tilde{f}(r) \to 1$ at spatial infinity.
This allows us to realize the asymptotically Minkowski background.

Around the outer horizon characterized by the radius $r_h$, we expand the solutions in the form 
\begin{equation}
f=\sum_{i=1} f_i \left( r/r_h-1 \right)^i\,,\qquad 
h=\sum_{i=1} h_i \left(r/r_h-1 \right)^i\,,\qquad
\frac{A_0}{\Mpl}=a_0+\sum_{i=1} a_i \left(r/r_h-1 \right)^i\,,
\label{expanho}
\end{equation}
where $f_i$, $h_i$, $a_0$, and $a_i$ are dimensionless constants.
We assume that $f_1>0$ and $h_1>0$ to ensure the positivity of $f(r)$ and $h(r)$ outside the horizon. 
Taking the plus branch in Eq.~(\ref{back4d}), we have 
\begin{equation}
\frac{A_1}{\Mpl}=\epsilon\left[\frac{a_0}{\sqrt{f_1 h_1}} 
\left( r/r_h-1 \right)^{-1}
-\frac{a_0[f_1 (4h_1+h_2)+h_1 f_2]}
{2(f_1 h_1)^{3/2}}+{\cal O} (r/r_h-1)\right]\,,
\end{equation}
where $\epsilon=+1$ for $a_0>0$ and $\epsilon=-1$ for $a_0<0$. 
Hereafter, we set $\epsilon=+1$ without loss of generality since the Lagrangian \eqref{action} is invariant under the transformation $A_{\mu}\to-A_{\mu}$ and $\beta_3\to-\beta_3$. 
Using the three coefficients $f_1$, $a_0$, and $a_1$, we can express the other coefficients in the expansions (\ref{expanho}).
For example, we have
\begin{eqnarray}
h_1 &=& \frac{2f_1}{2f_1+a_1^2}\,,\label{h1b}\\
f_2 &=&[2f_1^4 a_1^3 - 4 f_1^5 a_1 - b_3^2 a_0(2 a_0 + a_1) 
(2f_1+a_1^2 )(8 a_0 a_1 + 3 a_1^2 + 2 f_1) 
\{40 f_1 a_0^3 - f_1 a_1^3 + 2a_0^2 (7 f_1 a_1+a_1^3 )  \nonumber \\
& &
+ a_0 (a_1^4 - 3  f_1 a_1^2 - 2 f_1^2) \} 
- b_3 (4 a_0 + a_1) f_1^2 \{ 2 a_0 a_1^3 (2a_0 + a_1) 
+ f_1a_1(a_1^2-18a_0^2 - 5a_0 a_1 ) - 2 f_1^2a_0  \} \nonumber \\
& &
\times
\sqrt{4f_1+2a_1^2}]/[ \sqrt{2}f_1^2 {\cal K} 
\{ 2b_3a_0(8a_0a_1 + 3a_1^2+2f_1 )(2a_0 + a_1) 
\sqrt{2f_1+a_1^2} - 2\sqrt{2} f_1^2 a_1 \}]\,,\\ 
h_2 &=& [\{8a_0^3 (f_1+2a_1^2)+ 2a_0^2 (5f_1 a_1+7 a_1^3) 
+ a_0 ( 2 f_1-3 a_1^2) ( f_1-a_1^2)-3 f_1 a_1^3\} 
b_3\sqrt{2f_1+a_1^2} \nonumber \\
& &
+\sqrt{2}f_1^2 ( 2f_1-a_1^2)]/
[(2f_1+a_1^2) \sqrt{2}f_1^2 {\cal K}]\,,
\label{h2b}
\end{eqnarray}
where $b_3$ and ${\cal K}$ are dimensionless quantities defined, respectively, by 
\ba
b_3 &\equiv& r_h \Mpl \beta_3\,,\\
{\cal K} &\equiv& \frac{b_3 a_0[4 a_0 (2 a_0 + a_1) - f_1]
\sqrt{a_1^2 + 2 f_1}}
{\sqrt{2} f_1^2}-1\,.
\label{calK}
\ea
Note that an analogous relation can be also written for the term $a_2$. 
As we will see in Sec.~\ref{sta1sec}, the quantity ${\cal K}$ is related to the linear stability condition on the horizon. 
To avoid the Laplacian instability of one of the dynamical perturbations, we will show 
that the inequality,
\be
{\cal K}>0\,,
\label{Kcon}
\ee
needs to hold. Under this condition, the divergences of $f_2$ and $h_2$ at ${\cal K}=0$ can be avoided. 
In the limit that $b_3 \to 0$, we have ${\cal K} \to -1$ and hence the condition (\ref{Kcon}) is violated.

For a given value of $b_3$, we numerically integrate Eqs.~(\ref{back1})-(\ref{back3}) with Eq.~(\ref{back4}) outwards by using the boundary conditions (\ref{expanho}) around $r=r_h$. 
In the first run, we choose $f_1=1$ and obtain a numerical value of $f_{\infty}$ by integrating the background equations up to a sufficiently large distance. 
For the second integration, we perform the following redefinitions 
\be
f_1=1/f_{\infty}\,,\qquad
a_0=a_0^{\rm first}/\sqrt{f_{\infty}}\,,\qquad 
a_1=a_1^{\rm first}/\sqrt{f_{\infty}}\,,
\label{rescaling}
\ee
where the upper subscript ``first'' means the values of $a_0$ and $a_1$ in the first run. 
Then, the metric component $f$ asymptotically approaches 1 in the second run.

\begin{figure}[ht]
\begin{center}
\includegraphics[height=3.5in,width=3.5in]{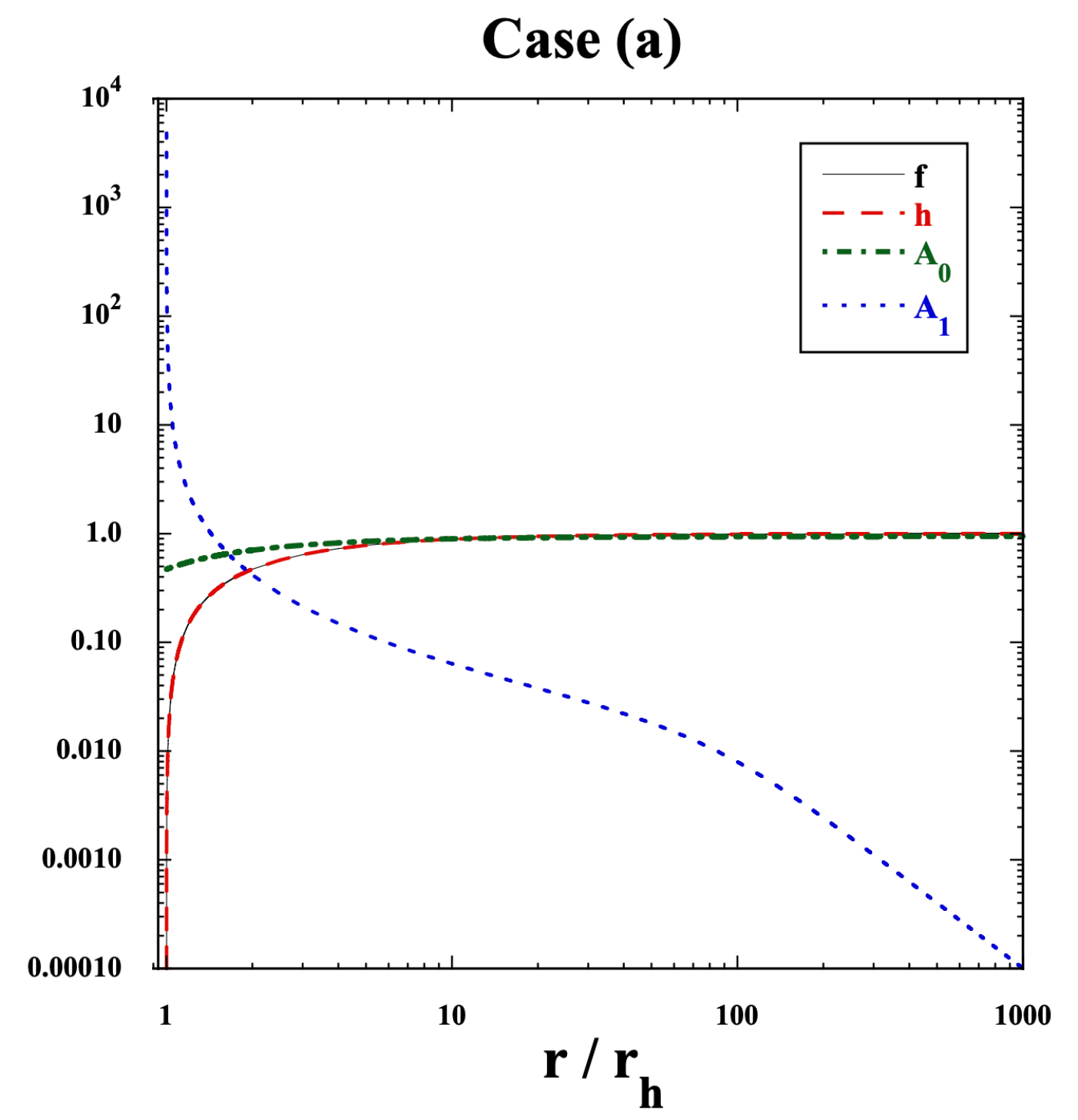}
\includegraphics[height=3.5in,width=3.5in]{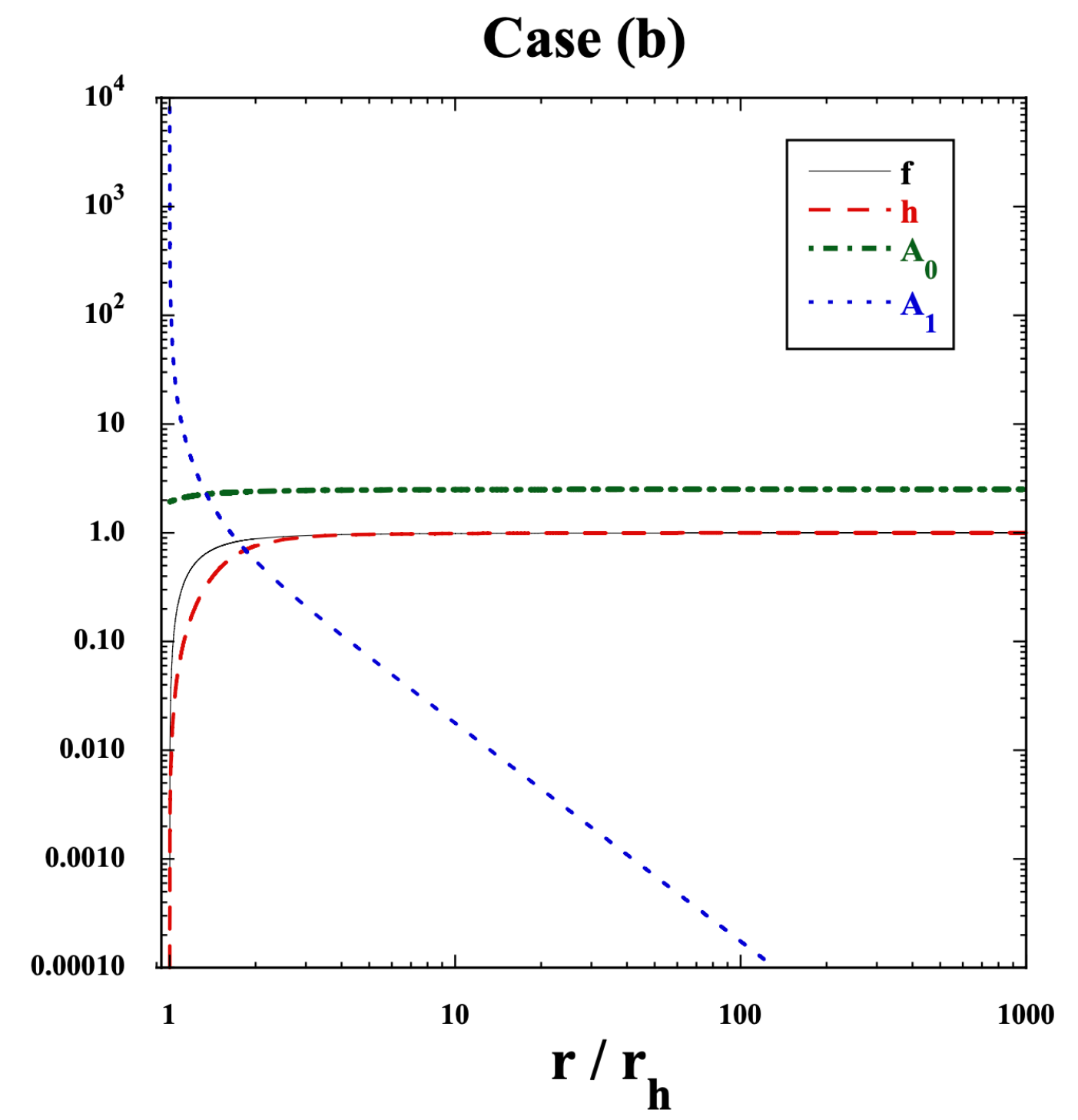}
\end{center}\vspace{-0.5cm}
\caption{\label{fig1}
We plot $f$, $h$, $A_0$, and $A_1$ versus $r/r_h$ for two different cubic couplings $b_3$.
The case (a) corresponds to $b_3=1.0 \times 10^{-3}$ with $f_1=0.8861$, $a_0=0.4707$, and $a_1=0.4707$, whereas the case (b) corresponds to $b_3=1.0$ with $f_1=3.7380$, $a_0=1.9334$, and $a_1=1.9334$.}
\end{figure}

In Fig.~\ref{fig1}, we show the radial dependence of $f$, $h$, $A_0$, and $A_1$ outside the horizon for two different cubic couplings: (a) $b_3=1.0 \times 10^{-3}$ (left) and (b) $b_3=1.0$ (right). 
The boundary conditions at $r=r_h$, i.e., $f_1$, $a_0$, and $a_1$, are chosen such that $f$ asymptotically approaches 1 according to the prescription explained above. We note that the hairy BH solution for the small coupling, e.g., case (a), was also studied in Ref.~\cite{Heisenberg:2017xda}, 
but the issue of rescalings (\ref{rescaling}) was not addressed there. Indeed, the proper rescaling is particularly important for large $b_3$ like case (b), as the deviation of $f_{\infty}$ from 1 in the first run tends to be significant.

In case (a), which corresponds to a small coupling $b_3$ much less than 1, the quantity ${\cal K}$ defined in Eq.~(\ref{calK}) is negative. 
Hence this case is excluded by the instability of perturbations, but we will discuss the property of background BH solutions to make a comparison with the large coupling case. 
As we see in the left panel of Fig.~\ref{fig1}, the radial dependence of $f$ is similar to that 
of $h$ throughout the horizon exterior. 
We observe that both $f$ and $h$ asymptotically approach 1 at large distances. 
The temporal vector component varies as $A_0 \simeq \left[ a_0+a_1(r/r_h-1) \right]\Mpl$ around the horizon and it approaches a constant value $P \simeq 0.9448 \Mpl$ as $r \to \infty$. 
In case (a), which corresponds to $a_0=a_1=0.4707$, the asymptotic value of $A_0$ can be estimated as 
$P \simeq (a_0+a_1)\Mpl$. For $b_3 \ll 1$, $P$ is typically smaller than the order $\Mpl$ by reflecting the property that neither $a_0$ nor $a_1$ exceeds 1. 
The longitudinal vector component exhibits the divergence $A_1 \simeq \Mpl(a_0/\sqrt{f_1 h_1}) (r/r_h-1)^{-1}$ at $r=r_h$, but the scalar product $X$ remains finite. 
In the left panel of Fig.~\ref{fig1}, we can confirm that $A_1$ decreases in proportion to $q_s/r^2$ for the distance $r \gtrsim 100 r_h$.

In case (b), we have chosen $f_1^{\rm first}=1$, $a_0^{\rm first}=1$, and $a_1^{\rm first}=1$ 
for the first run and then obtained the value $f=0.26753$ at the distance $r=10^3 r_h$.
Then, we performed the second integration by setting $f_{\infty}=0.26753$ in Eq.~(\ref{rescaling}), 
which results in the values of $f_1=3.7379$, $a_0=1.9334$, and $a_1=1.9334$.
In case (b), we have ${\cal K}>0$, and hence the Laplacian instability is absent on the horizon. 
Let us consider positive values of $b_3$, $a_0$, and $a_1$ satisfying the condition (\ref{Kcon}). 
At least, we require that the term $4a_0(2a_0+a_1)-f_1$ in ${\cal K}$ is positive, so that 
\be
a_0>\frac{\sqrt{a_1^2+2f_1}-a_1}{4}\,.
\label{a0bound}
\ee
In case (b), this inequality translates to the bound $a_0>0.3538$.
For the parameter range in which the condition ${\cal K}>0$ is satisfied, we generally have 
$f_1 \gtrsim 1$ to realize the asymptotic behavior $f(r \to \infty)=1$. 
When $a_1 \lesssim f_1$, the inequality (\ref{a0bound}) gives the bound $a_0>{\cal O}(0.1)$.
Since the limit (\ref{a0bound}) is merely a necessary condition for the realization of ${\cal K}>0$, 
we typically require tighter bounds $a_0 \gtrsim {\cal O}(1)$ for $a_1 \lesssim f_1$, 
as it happens in case (b).
When $a_1 \gtrsim f_1={\cal O}(1)$, the lower limit of $a_0$ in Eq.~(\ref{a0bound}) can be smaller than the order 0.1, but $a_1$ exceeds the order of 1. 
In both cases of $a_1 \lesssim f_1$ and $a_1 \gtrsim f_1$, either $a_0$ or $a_1$ is larger than the order 1, so that 
\be
a_0 + a_1 \gtrsim {\cal O}(1)\,.
\label{a01}
\ee
For such boundary conditions of $a_0$ and $a_1$, the general behavior of solutions is that $A_0$ 
increases from the value $a_0 \Mpl$ on the horizon toward an asymptotic value $P$ exceeding the order of $\Mpl$. 
In case (b), we have $A_0=2.5132\Mpl$ at $r=10^3 r_h$.
As we will see in Sec.~\ref{sta1sec}, the asymptotic value of $A_0$ in the range $|P|>\sqrt{2} \Mpl$ 
leads to the angular Laplacian instability of perturbations far away from the horizon. 
For $b_3$, $f_1$, $a_0$, and $a_1$ consistent with the condition (\ref{Kcon}), we generally have the 
asymptotic solution to $A_0$ whose magnitude is in the region $|P|>\sqrt{2}\Mpl$.
 
In the right panel of Fig.~\ref{fig1}, we observe that there is a difference between $f$ and $h$ around the horizon and that both $f$ and $h$ converge to 1 in the regime $r \gg r_h$. 
The former property is different from that of the case $b_3 \ll 1$ in which $f$ and $h$ are similar to each other throughout the horizon exterior.
In comparison to case (a), the longitudinal component in case (b) enters the region with $A_1 \simeq q_s/r^2$ at a smaller distance $r$ outside the horizon.

\subsection{$m \neq 0$ and $\beta_3 \neq 0$}
\label{nohairBHsec}

Let us discuss the background BH solutions for $m \neq 0$ and $\beta_3 \neq 0$.  
At spatial infinity, we expand $f$, $h$, $A_0$, and $A_1$ in the forms (\ref{largeex}) and substitute them into Eqs.~(\ref{back1})-(\ref{back4}). 
Then, it follows that all the coefficients in $A_0$ and $A_1$ vanish, i.e., 
\begin{equation}
P=0\,,\qquad \tilde{a}_i=0\,,\qquad 
\tilde{b}_0=0\,,\qquad \tilde{b}_i=0\,,\qquad 
{\rm for} \quad i \geq 1\,.   
\end{equation}
Moreover, we obtain 
\begin{equation}
\tilde{f}_1=f_{\infty} \tilde{h}_1\,,\qquad h_{\infty}=1\,,\qquad 
\tilde{f}_{i}=\tilde{h}_{i}=0\,,\qquad {\rm for} \quad i \geq 2\,,
\end{equation}
where we will set $f_{\infty}=1$ and $\tilde{h}_1=-2M$ in the following. 
This corresponds to the no-hair Schwarzschild solution 
\begin{equation}
f=h=1-\frac{2M}{r}\,,\qquad 
A_0=0\,,\qquad A_1=0\,,
\label{Sch}
\end{equation}
for the distance $r \gg r_h$. 

Let us also consider the behavior of solutions around the outer horizon by using the expansions (\ref{expanho}). 
At $r=r_h$, the coefficient of $A_0'$ in Eq.~(\ref{back3}) takes a finite value 
\begin{equation}
p_1 \equiv r_h \left( \frac{2}{r}-\frac{f'}{2f}+\frac{h'}{2h} \right) \Biggr|_{r=r_h}
=\frac{f_1 (4h_1+h_2)-h_1 f_2}{2f_1 h_1}\,.
\end{equation}
Around $r=r_h$, the longitudinal vector component has the following behavior
\begin{equation}
\frac{A_1}{\Mpl}=\frac{a_0}{\sqrt{f_1 h_1}} 
\left( r/r_h-1 \right)^{-1}
-\frac{a_0 b_3[f_1 (4h_1+h_2)+h_1 f_2]
+2\bar{m}^2  f_1^{3/2} \sqrt{h_1}}
{2b_3 (f_1 h_1)^{3/2}}+{\cal O} (r/r_h-1)\,,
\end{equation}
where 
\begin{equation}
\bar{m} \equiv m r_h\,.
\end{equation}
The mass-dependent coefficient $-m^2/h$ in Eq.~(\ref{back3}) is divergent on the horizon, but the $ A_1$-dependent terms cancel its divergence. 
Indeed, we have 
\begin{eqnarray}
p_2 &\equiv& -r_h^2 \left[ \frac{m^2}{h}+\beta_3 \left\{  A_1'+\left( 
\frac{2}{r}+\frac{f'}{2f}+\frac{h'}{2h} \right)A_1 \right\} \right] \Biggr|_{r=r_h}\nonumber \\
&=& -\frac{2 f_1^2 h_1 \bar{m}^4 + \sqrt{f_1 h_1} a_0 b_3 \bar{m}^2 
(12f_1h_1 + f_1 h_2 + 3 h_1 f_2) + 2 b_3^2 h_1^2  
(10 a_0^2 f_1 + 4 a_0^2 f_2 + 2 a_0 a_1 f_2 - 2 a_0 a_ 2 f_1 - a_1^2 f_1)}
{2 f_1^{3/2} h_1^{5/2} a_0 b_3}\,,\nonumber \\
\end{eqnarray}
which is constant.

Solving the differential equation $A_0''+(p_1/r_h)A_0'+(p_2/r_h^2) A_0=0$ for $A_0$ around $r=r_h$, we obtain 
\begin{equation}
A_0=C_1 \exp \left[\frac{(-p_1+\sqrt{p_1^2-4p_2})r}{2r_h} \right]
+C_2 \exp \left[\frac{(-p_1-\sqrt{p_1^2-4p_2})r}{2r_h} \right]\,.
\label{A0}
\end{equation}
For nonvanishing integration constants $C_1$ and $C_2$, there is a variation of $A_0$ in the vicinity of the horizon.  
In this case, the temporal vector component does not generally vanish as the distance $r$ increases from $r_h$.  
In the region where $f$ and $h$ start to approach 1, the equation for $A_0$ contains a growing-mode solution $A_0 \propto e^{mr}/r$, whose dominance contradicts with the asymptotic solution (\ref{Sch}).
To avoid such a behavior, we need to choose $C_1=0=C_2$ in Eq.~(\ref{A0}), in which case $a_0=0=a_i$ 
for all $i \geq 1$ with $A_1(r_h)=0$. 
Thus, in the two asymptotic regimes, we require that both $A_0$ and $A_1$ are vanishing.
Even in the intermediate regime, we need to satisfy the conditions $A_0(r)=0$ and $A_1(r)=0$ to avoid 
the appearance of the aforementioned growing mode. 
Hence we end up with the no-hair Schwarzschild solution (\ref{Sch}) throughout the horizon exterior. 
Thus, the introduction of the mass term $m$ to cubic vector Galileons removes the existence 
of hairy BH solutions.

\section{Black hole perturbations}
\label{gaugesec}

In this section, we perform the formulation of BH perturbations in GP theories and discuss the gauge choices to study the linear stability of BHs in the presence of cubic vector Galileons. 
Since the perturbations in the odd-parity sector were already studied in Ref.~\cite{Kase:2018voo}, we will briefly revisit them at the end of this section.
In Secs.~\ref{sta1sec} and \ref{sta2sec}, we will derive linear stability conditions of even-parity perturbations for the BH solutions presented in Secs.~\ref{hairBHsec} and \ref{nohairBHsec}, respectively, by choosing particular gauge conditions. 
For the perturbations on the SSS background (\ref{metric}), we can choose the $m=0$ component of spherical harmonics $Y_{lm}(\theta,\varphi)$ without loss of generality.  
Metric perturbations have the following components \cite{Regge:1957td, Zerilli:1970se} 
\begin{eqnarray}
& &
h_{tt}=f(r) H_0 (t,r) Y_{l} (\theta)\,,\qquad 
h_{t r}=h_{r t}=H_1 (t,r) Y_{l} (\theta)\,,\qquad 
h_{t \theta}=h_{\theta t}=h_0 (t,r)
Y_{l,\theta}(\theta)\,,\nonumber \\
& &
h_{t \varphi}=h_{\varphi t}=-Q (t,r) 
(\sin \theta)Y_{l,\theta}(\theta)\,,\qquad 
h_{rr}=h^{-1}(r) H_2(t,r)Y_{l}(\theta)\,,\qquad 
h_{r \theta}=h_{\theta r}=h_1 (t,r) 
Y_{l,\theta}(\theta)\,,\nonumber \\
& &
h_{r \varphi}=h_{\varphi r}=-W (t,r) 
(\sin \theta)Y_{l,\theta}(\theta)\,,\qquad 
h_{\theta \theta}=r^2 K(t,r)Y_{l} (\theta)
+r^2 G(t,r)Y_{l,\theta \theta} (\theta)\,,
\nonumber \\
& &
h_{\varphi \varphi}=r^2 K(t,r)(\sin^2 \theta) 
Y_{l} (\theta)+r^2 G(t,r)
(\sin \theta)(\cos \theta)Y_{l,\theta} (\theta)\,,
\nonumber\\
& &
h_{\theta \varphi}=\frac{1}{2}U(t,r) 
\left[ (\cos \theta)Y_{l,\theta}(\theta)
-(\sin \theta)Y_{l,\theta \theta} (\theta) 
\right]\,,
\end{eqnarray}
where we have used the notations $Y_l(\theta) \equiv Y_{l0} (\theta)$, $Y_{l,\theta} \equiv \rd Y_l(\theta)/\rd \theta$, and $Y_{l,\theta \theta} \equiv \rd^2 Y_l(\theta)/\rd \theta^2$, and the summation of $Y_{l}(\theta)$ with respect to the multiples $l$ is omitted. 
Since we are interested in the stability of BHs in the eikonal limit, we will focus on the multipoles $l \geq 2$ in the following discussion.
The odd-parity modes, which are characterized by the parity $(-1)^{l+1}$, correspond to three metric 
perturbations $Q$, $W$, and $U$. 
In the even-parity sector with the parity $(-1)^l$, there are seven perturbations $H_0$, $H_1$, $h_0$, $H_2$, $h_1$, $K$, and $G$.
The vector field $A_{\mu}$ has the following perturbed components \cite{Moncrief:1974gw, Moncrief:1974ng, Zerilli:1974ai, Moncrief:1975sb}
\begin{equation}
\delta A_{t}= \delta A_0 (t,r) Y_{l}(\theta)\,,\qquad 
\delta A_{r}= \delta A_1 (t,r) Y_{l}(\theta)\,,\qquad
\delta A_{\theta}= \delta A_2 (t,r) 
Y_{l,\theta}(\theta)\,,\qquad
\delta A_{\varphi}= -\delta A (t,r)
(\sin \theta)Y_{l,\theta}(\theta)\,,
\label{delu}
\end{equation}
with the background value (\ref{vecback}). 
The three components $\delta A_0$, $\delta A_1$, and $\delta A_2$ belong to the perturbations in the even-parity sector, whereas $\delta A$ corresponds to the odd-parity mode. 

Let us consider the following infinitesimal gauge transformation
\begin{equation}
\tilde{x}_{\mu}=x_{\mu}+\xi_{\mu}\,,
\label{gatrans}
\end{equation}
where $\xi_{\mu}$ has the following components
\begin{equation}
\xi_{t}={\cal T}(t,r)Y_{l} (\theta)\,,\qquad
\xi_{r}={\cal R}(t,r)Y_{l} (\theta)\,,\qquad
\xi_{\theta}=\Theta (t,r) Y_{l,\theta} (\theta)\,,\qquad
\xi_{\varphi}=-\Lambda (t,r) (\sin \theta) 
Y_{l,\theta} (\theta)\,.
\end{equation}
At linear order in $\xi_{\mu}$, the odd-parity metric perturbations transform as 
\begin{equation}
\tilde{Q}= 
Q-\dot{\Lambda}\,,\qquad
\tilde{W}= 
W-\Lambda'+\frac{2}{r}\Lambda\,,\qquad
\tilde{U}=U-2\Lambda\,,
\end{equation}
where a dot represents the derivative with respect to $t$.
The transformation law for even-parity metric perturbations is given by 
\begin{eqnarray}
& &
\tilde{H}_0=
H_0-\frac{2}{f} \dot{\cal T}+\frac{f' h}{f}{\cal R}\,,\qquad 
\tilde{H}_1=
H_1-\dot{{\cal R}}-{\cal T}'+\frac{f'}{f}{\cal T}\,,\qquad 
\tilde{H}_2 = H_2-2h{\cal R}'-h' {\cal R}\,,\nonumber \\
& &
\tilde{h}_0= h_0-{\cal T}-\dot{\Theta}\,,\qquad 
\tilde{h}_1= h_1-{\cal R}-\Theta'+\frac{2}{r} \Theta\,,
\qquad 
\tilde{K} = K-\frac{2}{r}h {\cal R}\,,\qquad
\tilde{G} = G-\frac{2\Theta}{r^2}\,.
\end{eqnarray}
The vector-field perturbation $\delta A_{\mu}$ is subject to the transformation 
\begin{equation}
\widetilde{\delta A}_{\mu}=\delta A_{\mu} 
-A_{\alpha}\,{\xi^{\alpha}}_{,\mu}
-A_{\mu,\alpha} \xi^{\alpha}\,,
\label{Leedu}
\end{equation}
where $A_{\mu}$ is the background value given by Eq.~(\ref{vecback}). 
Then, the components of $\delta A_{\mu}$ transform, respectively, as 
\begin{eqnarray}
\widetilde{\delta A}_{0} &=& 
\delta A_0+\frac{A_0}{f}\dot{{\cal T}} 
-h A_1 \dot{\cal R}-h A_0' {\cal R}\,,\\
\widetilde{\delta A}_{1} &=& 
\delta A_1+\frac{A_0}{f}{\cal T}'
-\frac{f' A_0}{f^2}{\cal T}-h A_1 {\cal R} 
-\left( h' A_1+h A_1' \right){\cal R}\,,\\
\widetilde{\delta A}_{2} &=& 
\delta A_2+\frac{A_0}{f}{\cal T}-h A_1 {\cal R}\,,\\
\widetilde{\delta A} &=&\delta A\,.
\end{eqnarray}

We can choose gauge conditions to fix the four components ${\cal T}$, ${\cal R}$, $\Theta$, and $\Lambda$. In the odd-parity sector, the typical gauge choice is given by \cite{Moncrief:1974ng, Kobayashi:2012kh, Kase:2018voo, Kase:2023kvq}
\begin{equation}
\tilde{U}=0\,,
\end{equation}
which fixes $\Lambda$ to be $\Lambda=U/2$.
Then, the three fields $\tilde{Q}$, $\tilde{W}$, and $\widetilde{\delta A}$ are left for the analysis 
of odd-parity perturbations.  
We can define a gravitational dynamical perturbation $\tilde{\chi}$ constructed from $\tilde{Q}$ and 
$\tilde{W}$ \cite{Kase:2018voo}, so that two propagating DOFs are present in the odd-parity sector.

In the even-parity sector, we have several possible gauge choices. One of them is given by 
\begin{equation}
\tilde{G}=0\,,\qquad \tilde{h}_1=0\,,\qquad 
\widetilde{\delta A}_{2}=0\,,
\label{gauge1}
\end{equation}
under which ${\cal T}$, ${\cal R}$, and $\Theta$ are fixed. 
A similar gauge was chosen in an Aether-orthogonal frame in Einstein-Aether theories \cite{DeFelice:2024dbj}, after which there are seven fields $\tilde{H}_0$, $\tilde{H}_1$, $\tilde{h}_0$, $\widetilde{\delta A}_{0}$, $\tilde{H}_2$, $\widetilde{\delta A}_{1}$, and $\tilde{K}$ 
left in the analysis. 
In Einstein-Aether theories, eliminating the nondynamical perturbations $\tilde{H}_0$, $\tilde{H}_1$, $\tilde{h}_0$, and $\widetilde{\delta A}_{0}$ ends up with three propagating DOFs $\tilde{H}_2$, $\widetilde{\delta A}_{1}$, and $\tilde{K}$, which correspond to the dynamical perturbations in the tensor, vector, and scalar sectors, respectively. 

In GP theories we can potentially choose the gauge conditions (\ref{gauge1}) for even-parity perturbations. 
For the gauge choice (\ref{gauge1}), however, some higher-order derivative terms like $K''^2$ appear after removing the field $H_0$ from the action. Such terms are attributed to the combination $L\tilde{G}-2\tilde{K}$ and their derivatives in the second-order action, where  
\begin{equation}
L \equiv l(l+1)\,.
\end{equation}
To avoid the appearance of those terms, we may choose the gauge condition $\tilde{G}=2\tilde{K}/L$ instead of $\tilde{G}=0$, so that the higher-order derivatives of $\tilde{K}$ can be eliminated from the action at the same time as the derivatives of $\tilde{G}$.
In GP theories where the background values of $A_0$ and $A_1$ are nonvanishing, the gauge choice
\begin{equation}
L\tilde{G}=2\tilde{K}\,,\qquad \tilde{h}_1=0\,,\qquad 
\widetilde{\delta A}_{2}=0\,,
\label{gauge2}
\end{equation}
can be at work. 
In Sec.~\ref{sta1sec}, we will exploit this gauge condition to derive the linear stability conditions of hairy BHs with $A_0 \neq 0$ and $A_1 \neq 0$ present in theories with $m=0$ and $\beta_3 \neq 0$.

In the case of no-hair BH solutions where both $A_0$ and $A_1$ are vanishing, we need to choose gauges different from (\ref{gauge1}) or (\ref{gauge2}). 
One of the possible choices is given by 
\begin{equation}
\tilde{G}=0\,,\qquad \tilde{h}_1=0\,,\qquad 
\tilde{h}_0=0\,,
\label{gauge3}
\end{equation}
instead of setting $\widetilde{\delta A}_{2}=0$. 
In Sec.~\ref{sta2sec}, we will apply this gauge choice to the linear stability of no-hair BHs present for $m \neq 0$ and $\beta_3 \neq 0$.
Note that none of the three gauge conditions given above is the same as the gauge chosen in Refs.~\cite{Kobayashi:2014wsa} in the context of Horndeski theories or theories with $U(1)$ gauge invariance \cite{Kase:2023kvq, DeFelice:2024eoj, Chen:2024hkm, Zhang:2024cbw}.
In the following, we omit the tilde from the gauge-transformed fields.

Before moving on to the analysis of even-parity perturbations, we summarise the linear stability 
conditions against odd-parity perturbations in the small-scale limit. Note that they were already 
derived in Ref.~\cite{Kase:2018voo} for full GP theories. 
We are now considering theories with the action (\ref{action}), which amount to the couplings $G_2=m^2 X$, $G_3=\beta_3 X$, and $G_4=\Mpl^2/2$ in the notation of Ref.~\cite{Kase:2018voo}. 
Two dynamical perturbations in the odd-parity sector are given by 
\be
\chi \equiv 
\dot{W}-Q'+\frac{2}{r}Q-\frac{2A_0'}{\Mpl^2} \delta A\,,
\quad {\rm and} \quad
\delta A\,, 
\ee
which correspond to the gravitational and vector-field perturbations, respectively. 
The ghosts are absent under the two conditions $q_1>0$ and $q_2>0$, where $q_1$ and $q_2$ are defined 
in Eqs.~(3.23) and (3.24) of Ref.~\cite{Kase:2018voo}. 
In current theories, they reduce, respectively, to 
\be
q_1=\frac{\Mpl^2 h}{4f^2}\,,\qquad 
q_2=\frac{1}{2r^2f}\,,
\ee
which are both positive outside the outer horizon ($f>0$ and $h>0$). 
The propagation speeds of $\chi$ and $\delta A$ along the radial direction are given by Eqs.~(3.28) 
and (3.29) in Ref.~\cite{Kase:2018voo}.
In current theories, we have 
\be
c_{r1}^2=1\,,\qquad 
c_{r2}^2=1\,,
\ee
which are both luminal.
The squared angular propagation speeds of two dynamical perturbations are expressed as Eq.~(3.37) in Ref.~\cite{Kase:2018voo}.
In current theories, they become 
\be
c_{\Omega+}^2=1\,,\qquad 
c_{\Omega-}^2=1\,.
\ee
Thus, the dynamical perturbations $\chi$ and $\delta A$ propagate with the speed of light both along the radial and angular directions. 
Hence there are neither ghost nor Laplacian instabilities in the odd-parity sector for theories given by the action (\ref{action}).

\section{Stability of BHs for $m=0$ and $\beta_3 \neq 0$}
\label{sta1sec}

As we discussed in Sec.~\ref{hairBHsec}, there are hairy BHs present in theories with $m=0$ and $\beta_3 \neq 0$. 
In this case, the temporal and longitudinal vector components are nonvanishing outside the outer horizon characterized by the radius $r_h$. 
To study the stability of hairy BHs against even-parity perturbations, we choose the gauge condition 
(\ref{gauge2}), i.e., 
\be
L G=2K\,,\qquad h_1=0\,,\qquad 
\delta A_{2}=0\,,
\label{gauge2d}
\ee
under which the seven fields $H_0$, $H_1$, $h_0$, $\delta A_{0}$, $H_2$, $K$, and $\delta A_{1}$ are left in the analysis.
We expand the action (\ref{action}) up to quadratic order in even-parity perturbations and perform the integration with respect to $\theta$ and $\varphi$. 
The resulting second-order action contains the products of the perturbed fields and their derivatives, with $r$-dependent coefficients.
Unlike Horndeski theories \cite{Kobayashi:2014wsa}, $U(1)$ gauge-invariant Maxwell-Horndeski theories \cite{Kase:2023kvq}, and $U(1)$ gauge-invariant scalar-vector-tensor theories \cite{Zhang:2024cbw}, 
the quadratic perturbed action contains the product $H_0^2$ besides terms linear in $H_0$. In this case, the perturbation equation for $H_0$ gives a constraint on the field $H_0$ itself.

We vary the second-order action with respect to the seven fields $H_0$, $H_1$, $h_0$, $\delta A_{0}$, 
$H_2$, $K$, and $\delta A_{1}$.  
To discuss the stability of BHs in the small-scale limit, we assume the solutions to the perturbation equations in the form 
\ba
& &
H_0=\tilde{H}_0 e^{-i (\omega t-kr)}\,,\qquad 
H_1=\tilde{H}_1 e^{-i (\omega t-kr)}\,,\qquad 
h_0=\tilde{h}_0 e^{-i (\omega t-kr)}\,,\qquad 
\delta A_0=\tilde{\delta A}_0 e^{-i (\omega t-kr)}\,,
\nonumber \\
& &
H_2=\tilde{H}_2 e^{-i (\omega t-kr)}\,,\qquad 
K=\tilde{K} e^{-i (\omega t-kr)}\,,\qquad 
\delta A_1=\tilde{\delta A}_1 e^{-i (\omega t-kr)}\,,
\label{WKB}
\ea
where $\tilde{H}_0$, $\tilde{H}_1$, $\tilde{h}_0$, $\tilde{\delta A}_0$, $\tilde{H}_2$, $\tilde{K}$, and $\tilde{\delta A}_1$ are constant under the WKB approximation. 
We are interested in the regime where the angular frequency $\omega$, the momentum $k$, and the multipole $l$ satisfy the conditions $\omega r_h \gg 1$, $k r_h \gg 1$, and $l \gg 1$. Substituting Eq.~(\ref{WKB}) into the perturbation equations of motion for $H_0$, $H_1$, $h_0$, and $\delta A_0$, 
we can express them in the following forms
\ba
& &
\alpha_1 \tilde{H}_0+\alpha_2 \tilde{H}_1
+\alpha_3 \tilde{h}_0
+\alpha_4 \tilde{\delta A}_0+\alpha_5 \tilde{H}_2
+\alpha_6 \tilde{K}+\alpha_7 \tilde{\delta A}_1=0\,,
\label{nondy1}\\
& &
\lambda_1 \tilde{H}_0+\lambda_2 \tilde{H}_1
+\lambda_3 \tilde{h}_0
+\lambda_4 \tilde{\delta A}_0+\lambda_5 \tilde{H}_2
+\lambda_6 \tilde{\delta A}_1=0\,,\\
& &
\mu_1 \tilde{H}_0+\mu_2 \tilde{H}_1+\mu_3 \tilde{h}_0
+\mu_4 \tilde{\delta A}_0+\mu_5 \tilde{H}_2
+\mu_6 \tilde{K}+\mu_7 \tilde{\delta A}_1=0\,,\\
& &
\nu_1 \tilde{H}_0+\nu_2 \tilde{H}_1+\nu_3 \tilde{h}_0
+\nu_4 \tilde{\delta A}_0+\nu_5 \tilde{H}_2
+\nu_6 \tilde{\delta A}_1=0\,,
\label{nondy4}
\ea
respectively, where $\alpha_{1,\cdots,7}$, $\lambda_{1,\cdots,6}$, $\mu_{1,\cdots,7}$, and $\nu_{1,\cdots,6}$ are $r$-dependent coefficients.
We solve Eqs.~(\ref{nondy1})-(\ref{nondy4}) for $\tilde{H}_0$, $\tilde{H}_1$, $\tilde{h}_0$, $\tilde{\delta A_0}$ and substitute them into the perturbation equations for $H_2$, $K$, and $\delta A_1$.
This process leads to the following algebraic equations
\be
{\bm V} \vec{{\cal X}}^{\rm T}=0\,,
\label{dyneq}
\ee
where ${\bm V}$ is a $3 \times 3$ matrix, and 
\be
\vec{{\cal X}}=\left( \tilde{H}_2, \tilde{K}, 
\tilde{\delta A}_1 
\right)\,.
\ee
Three equations of motion in Eq.~(\ref{dyneq}) correspond to those of the dynamical perturbations $H_2$, $K$, and $\delta A_1$. 
In the limit that $\omega r_h \gg 1$, $k r_h \gg 1$, and $l \gg 1$, the matrix ${\bm V}$ becomes Hermitian with their components satisfying the relation $V_{ij}=V_{ji}^*$.

\subsection{Linear stability conditions}

We can obtain the three no-ghost conditions by choosing terms proportional to $\omega^2$ in ${\bm V}$.
We express the matrix with these components as $\omega^2 {\bm V}^{\rm kin}$. 
To avoid ghosts, the determinants of submatrices of ${\bm V}^{\rm kin}$ are required to be positive.
The no-ghost condition of the vector-field perturbation $\delta A_1$ is  
\be
{\cal K}_1 \equiv V^{\rm kin}_{33}=\frac{r^2 L \sqrt{h}}
{(k^2 r^2 h+L) \sqrt{f}}>0\,,
\ee
which is automatically satisfied outside the outer horizon. 
The other two no-ghost conditions are given by 
\ba
{\cal K}_2 &\equiv& V^{\rm kin}_{11}V^{\rm kin}_{33}
-V^{\rm kin}_{13}V^{\rm kin}_{31} \nonumber \\
&=& L^3\Mpl^2 r^2 [4 \beta_3^2 r^2A_0^6 
-4 \beta_3^2 r^2 fh A_0^4 A_1^2+ \Mpl^2 f^3 
\{ 2\Mpl^2 (r h' + h - 1) -\beta_3 A_1^2 h(A_1 h' + 2A_1' h)
r^2 \} \nonumber \\
& &+ \Mpl^2 r f^2
\{ 2 \beta_3 rh A_0 A_0' A_1 + rh A_0'^2 
+ \beta_3 A_0^2 ( 3r h' A_1 + 6 rh A_1'+ 12h A_1
-2 \beta_3 rh A_1^2 )\}]  \nonumber \\
& &/[8\beta_3^2 f( k^2 r^2 h + L) A_0^2 
\{ A_0^2 L + h (A_0^2-A_1^2 f h) k^2 r^2\}^2]>0\,, 
\label{NG2}
\ea
and
\ba
{\cal K}_3 &\equiv& {\rm det}\,{\bm V}^{\rm kin} \nonumber \\
&=& L^3 \Mpl^4 r^4 [2 \Mpl^4 f^3 (h - 1) 
+ 2 f^2 \Mpl^2 (6 \beta_3 h A_0^2 A_1
+ \Mpl^2 f h')r + \{ \beta_3^2 A_0^2 (A_0^2-A_1^2 fh) 
(3 A_0^2+A_1^2 fh ) \nonumber \\
& &+ f^2 \{ 2 \beta_3 h A_0 A_0'A_1 
+ \beta_3 A_0^2 ( 3A_1 h'+ 6 A_1' h-2 \beta_3 h A_1^2) 
+ h [ A_0'^2 - \beta_3 A_1^2
f (A_1 h'+ 2 A_1' h)] \} \Mpl^2 \} r^2]\nonumber \\
& &
/[16 \beta_3^2 A_0^2 f^{3/2} \sqrt{h}(k^2 r^2 h + L)
\{  A_0^2 L + h(A_0^2-A_1^2 fh 
)k^2r^2 \}^2]>0\,.
\label{NG3}
\ea
The two conditions (\ref{NG2}) and (\ref{NG3}) need to be satisfied throughout the horizon exterior.

In terms of the proper time $\tau=\int \rd t\,\sqrt{f}$, the propagation speed $c_{\Omega}$ along the angular direction is defined by $c_{\Omega}=r \rd \theta/\rd \tau=(r/\sqrt{f})(\rd \theta/\rd t)
=(r/\sqrt{f})(\omega/l)$. 
To obtain the nonvanishing solutions to $\vec{{\cal X}}=( \tilde{H}_2, \tilde{K}, \tilde{\delta A}_1 )$, we require that the determinant of ${\bm V}$ is vanishing, i.e., ${\rm det}\,{\bm V}=0$.
Taking the limit $\omega r_h \approx l \gg k r_h \gg 1$ in this determinant equation, we obtain
\be
\left( c_{\Omega}^2-c_{\Omega,1}^2 \right)
\left( c_{\Omega}^2-1 \right)^2=0\,,
\ee
where 
\ba
c_{\Omega,1}^2
&=&
r \{ 2 f^2 \Mpl^2 [4 \beta_3 h A_0^2 A_1 + 
(f' h + h' f) \Mpl^2] - [2 \beta_3^2 A_0^2 
(A_0^2-A_1^2 fh)^2+2f \{ 
2 \beta_3 f h A_0 A_0' A_1-2 \beta_3^2 A_0^4 \nonumber \\
& &+ \beta_3 A_0^2 
(2 \beta_3 f h A_1^2-3 f' h A_1 -2 f h'A_1- 2f h A_1') 
+ f h [A_0'^2 + \beta_3 A_1^2 f (A_1 h' + 2A_1' h)] \}
\Mpl^2 \nonumber \\
& &
- ff'(f'h + h'f) \Mpl^4]r \}/
[4 \Mpl^4 f^3 (h - 1)+4\Mpl^2 f^2 (6 \beta_3 h A_0^2 A_1 
+ \Mpl^2 f h') r + 2\{ \beta_3^2 A_0^2 (A_0^2-A_1^2 f h)
(3 A_0^2+A_1^2 f h) \nonumber \\
& &+f^2 \Mpl^2 [2 \beta_3 h A_0 A_0' A_1 
+ \beta_3 A_0^2 ( 3 A_1 h' + 6 A_1' h-2\beta_3 h A_1^2) 
+ h A_0'^2 -\beta_3 A_1^2 fh (A_1 h' + 2A_1' h)] \} r^2]\,.
\label{cO1}
\ea

Then, we obtain the following three squared propagation speeds
\be
c_{\Omega}^2=c_{\Omega,1}^2\,,\qquad 
c_{\Omega}^2=1\,,\qquad c_{\Omega}^2=1\,,
\ee
While two of them are luminal, one of them has a nontrivial value. The angular Laplacian 
instability can be avoided for 
\be
c_{\Omega,1}^2>0\,.
\label{cOcon}
\ee

The propagation speed $c_r$ in the radial direction is defined in terms of the rescaled radial coordinate $r_*=\int \rd r/\sqrt{h}$ and the proper time $\tau=\int \rd t\,\sqrt{f}$, as 
$c_r=\rd r_*/\rd \tau=(fh)^{-1/2}(\rd r/\rd t)=(fh)^{-1/2}(\omega/k)$.
Taking the limit $\omega r_h \approx  k r_h \gg l \gg 1$, the determinant equation ${\rm det}\,{\bm V}=0$ leads to
\be
\left( c_{r}^2+s_1 c_r+s_2 \right)
\left( c_{r}^2-1 \right)^2=0\,,
\label{creq}
\ee
where $s_1$ and $s_2$ are functions of $r$ whose explicit forms are given in Appendix A. 
{}From Eq.~(\ref{creq}), we obtain the following solutions 
\be
c_r=\frac{-s_1 \pm 
\sqrt{s_1^2-4s_2}}{2}\,,\qquad 
c_r^2=1\,,\qquad c_r^2=1\,.
\ee
The latter two correspond to the luminal propagation speeds. 
On the other hand, since $s_1$ does not vanish in general, the first two solutions $c_r=(-s_1 \pm \sqrt{s_1^2-4s_2})/2$ have two different values of $c_r^2$.
If $s_1^2-4s_2$ is negative, the perturbations associated with one of the solutions to $c_r$ are subject to exponential instability. To avoid this behavior, we require that 
\be
s_1^2-4s_2>0\,,
\label{s12}
\ee
under which both of $c_r=(-s_1 \pm \sqrt{s_1^2-4s_2})/2$ are real. In this case, the perturbations exhibit oscillations without the exponential growth. 
However, the presence of two separate solutions to $c_r$ whose squared values are different from each other means that the propagation speed of one of the dynamical perturbations is not uniquely determined (for instance, if $s_1<0$ and $s_2>0$, we would have two positive solutions to $c_r$).
This shows an unhealthy property of hairy BHs present in theories with $m=0$ and $\beta_3 \neq 0$.
Indeed, we will show that the hairy BH does not generally satisfy all of the other linear stability conditions (\ref{NG2}), (\ref{NG3}), and (\ref{cOcon}).

\subsection{Application to hairy BHs}

Let us apply the above linear stability conditions to hairy BHs discussed in Sec.~\ref{hairBHsec}. 
On using the large-distance solutions (\ref{flar})-(\ref{fhA0}) far away from the outer horizon ($r \gg r_h$), the quantities ${\cal K}_2$ and ${\cal K}_3$ associated with the no-ghost conditions yield 
\ba
{\cal K}_2 &=&
\frac{\Mpl^2 L^3}{2 k^6 f_{\infty}}\frac{1}{r^2}+{\cal O} (r^{-3})\,,\\
{\cal K}_3 &=& \frac{3\Mpl^4 L^3}{16k^{6} {f_{\infty}}^{3/2}}+{\cal O} (r^{-1})\,.
\ea
Since we are considering the case $f_{\infty}=1$, the leading-order contributions to ${\cal K}_2$ 
and ${\cal K}_3$ are positive. 
Around $r=r_h$, we resort to the expanded solutions (\ref{expanho}) with Eqs.~(\ref{h1b})-(\ref{h1b}) for the estimation of ${\cal K}_2$ and ${\cal K}_3$. This gives
\ba
{\cal K}_2 &=& \frac{\sqrt{2}f_1 \Mpl^2 r_h^4}
{4\sqrt{a_1^2 + 2f_1} b_3 a_0^3}{\cal K}
+{\cal O} (r-r_h)\,,\label{calK2}\\
{\cal K}_3 &=& \frac{\Mpl^4 r_h^7}{8b_3 a_0^3 (r-r_h)}{\cal K}
+{\cal O} ((r-r_h)^0)\,,
\label{calK3}
\ea
where ${\cal K}$ is defined in Eq.~(\ref{calK}).
Outside the horizon, the leading-order contributions to Eqs.~(\ref{calK2}) and (\ref{calK3}) are positive for 
\be
b_3 a_0 {\cal K}>0\,.
\label{NGf}
\ee

In the regime $r \gg r_h$, the squared angular propagation speed (\ref{cO1}) can be estimated as
\be
c_{\Omega,1}^2=\frac{2\Mpl^2 f_{\infty}-P^2}{3P^2}
+{\cal O} (r^{-1})\,.
\label{cOL}
\ee
Then, the Laplacian instability is absent if
\be
2\Mpl^2 f_{\infty}>P^2\,.
\label{conL}
\ee
Around $r=r_h$, we have 
\be
c_{\Omega,1}^2 =\frac{2}{{\cal K}}+{\cal O}(r-r_h)\,,
\label{cOho}
\ee
and hence ${\cal K}>0$ to avoid the angular Laplacian instability. 
Combining ${\cal K}>0$ with the no-ghost condition (\ref{NGf}), it follows that 
\ba
\sqrt{2}f_1^2{\cal K}=
b_3 a_0[4 a_0 (2 a_0 + a_1) - f_1]
\sqrt{a_1^2 + 2 f_1}-\sqrt{2} f_1^2 &>&0\,,
\label{sta1}\\
b_3a_0 &>& 0\,.
\label{sta2}
\ea
For $b_3 \to 0$, the first inequality (\ref{sta1}) is violated. In this limit, we have 
$c_{\Omega,1}^2=-2$ at $r=r_h$, so there is the fatal Laplacian instability.
The condition (\ref{sta1}) requires large nonvanishing values of $b_3a_0$, in which case $c_{\Omega,1}^2$ is positive on the horizon.
We note that the condition (\ref{sta2}) demands that the signs of $b_3$ and $a_0$ are the same.
Without loss of generality, we will consider the case $b_3>0$ and $a_0>0$ in the following discussion.
Using the inequality (\ref{sta2}), it is at least necessary to satisfy the condition 
\be
4 a_0 (2 a_0 + a_1) >f_1\,,
\label{staf}
\ee
for the consistency with (\ref{sta1}). 
For positive $a_0$, the inequality (\ref{staf}) translates to (\ref{a0bound}).

\begin{figure}[ht]
\begin{center}
\includegraphics[height=3.5in,width=3.5in]{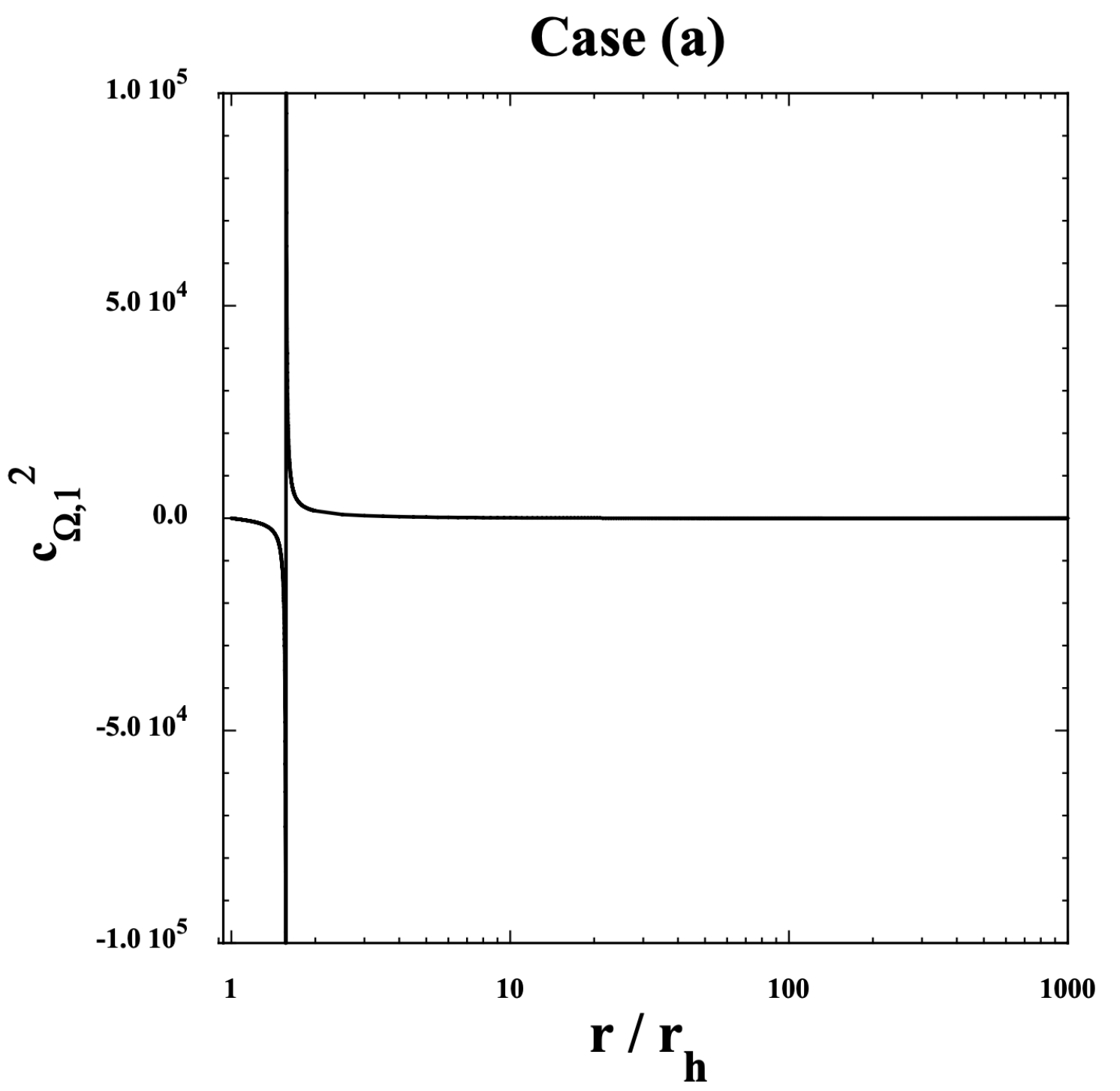}
\includegraphics[height=3.5in,width=3.5in]{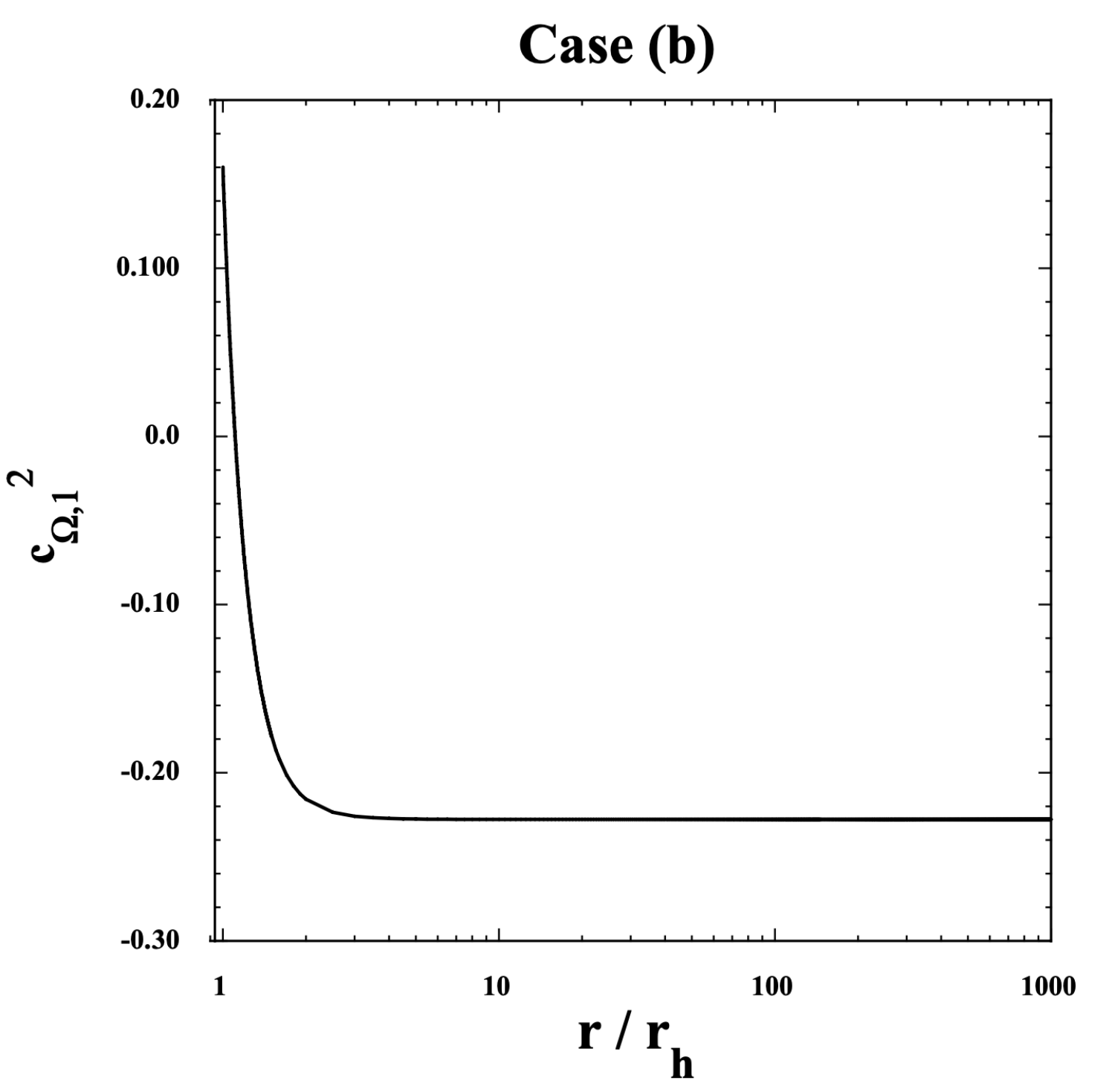}
\end{center}\vspace{-0.5cm}
\caption{\label{fig2}
The squared angular propagation speed $c_{\Omega,1}^2$ versus $r/r_h$, corresponding to the cases (a) and (b) in Fig.~\ref{fig1}. In case (a), $c_{\Omega,1}^2$ is close to $-2$ at $r=r_h$, with the asymptotic value $c_{\Omega,1}^2=0.41$ in the regime $r \gg r_h$.  
In case (b), we have $c_{\Omega,1}^2=0.16$ at $r=r_h$, with the asymptotic value $c_{\Omega,1}^2=-0.23$ for $r \gg r_h$. 
In both cases, the stability condition $c_{\Omega,1}^2>0$ is not satisfied in the whole region of the horizon exterior.}
\end{figure}

In the left panel of Fig.~\ref{fig2}, we plot $c_{\Omega,1}^2$ versus $r/r_h$ for the background BH solution corresponding to case (a) in Fig.~\ref{fig1}. 
Since the coupling $b_3$ is small in this case ($b_3=1.0 \times 10^{-3}$), the stability condition 
${\cal K}>0$, i.e., (\ref{sta1}), is violated. 
Indeed, $c_{\Omega,1}^2$ is close to $-2$ at $r=r_h$ and hence there is the Laplacian instability. 
Moreover, as estimated by Eqs.~(\ref{calK2}) and (\ref{calK3}) with ${\cal K}<0$, neither ${\cal K}_2>0$ nor ${\cal K}_3>0$ is satisfied in the vicinity of the horizon.
In the left panel of Fig.~\ref{fig2}, we observe that $c_{\Omega,1}^2$ exhibits the divergence at 
a finite distance outside the horizon.
This property is attributed to the fact that the denominator of $c_{\Omega,1}^2$ crosses 0 at an intermediate distance, which leads to the sign change of $c_{\Omega,1}^2$.
Far away from the horizon ($r \gg r_h$), we have $P=0.945\Mpl$ for $f_{\infty}=1$. 
In this case, the analytic estimate (\ref{cOL}) gives the asymptotic value $c_{\Omega,1}^2=0.41$, which is in good agreement with the numerical result.
We also note that both ${\cal K}_2$ and ${\cal K}_3$ are positive at large distances ($r \gg r_h$).
However, the violation of linear stability conditions on the horizon shows that the BH solution with 
small couplings $b_3$ in the range ${\cal K}<0$ is excluded. 

The right panel of Fig.~\ref{fig2} plots $c_{\Omega,1}^2$ versus $r/r_h$ corresponding to case (b) in Fig.~\ref{fig1}. 
Since $b_3$ is not small in this case ($b_3=1$), the linear stability conditions (\ref{sta1}) and (\ref{sta2}) are satisfied on the horizon, with $c_{\Omega,1}^2=0.16$ at $r=r_h$. 
However, we observe that $c_{\Omega,1}^2$ changes its sign at a distance $r_c$ outside the horizon, which is followed by the approach to the negative asymptotic value $c_{\Omega,1}^2 (r \gg r_h) \simeq -0.23$. 
This is in good agreement with the leading-order term of Eq.~(\ref{cOL}) derived by substituting 
the numerical value $P=2.513 \Mpl$. 
Since $c_{\Omega,1}^2<0$ for the distance $r>r_c$, the hairy BH in case (b) is prone to 
Laplacian instability. 

In addition to case (b), we studied the behavior of $c_{\Omega,1}^2$ for the coupling $b_3$ and boundary conditions of $f_1$, $a_0$, and $a_1$ satisfying the two stability conditions (\ref{sta1}) and (\ref{sta2}). 
As we discussed in Sec.~\ref{hairBHsec}, the existence of asymptotically Minkowski solutions consistent with the condition ${\cal K}>0$ on the horizon requires that $a_0$ and $a_1$ are in the ranges $a_0+a_1 \gtrsim {\cal O}(1)$.
We note that the choice of negative $a_1$ makes it harder to satisfy the inequality (\ref{sta1}).
We have run the numerical code by choosing many different combinations of $b_3$, $f_1$, $a_0$, and $a_1$ that realize the asymptotic behavior $f \to 1$ as $r \to \infty$. When the two inequalities 
(\ref{sta1}) and (\ref{sta2}) are satisfied, the general behavior of $c_{\Omega,1}^2$ is similar to the one shown in the right panel of Fig.~\ref{fig2}. 
Even if $c_{\Omega,1}^2>0$ on the horizon, $c_{\Omega,1}^2$ changes its sign at an intermediate 
distance $r_c$ and then it approaches a negative asymptotic value. 
The main reason for this behavior is that the realization of the condition (\ref{sta1}) along with $f(r \to \infty)=1$ requires that $a_0+a_1$ are greater than 1. 
Generally, this results in the asymptotic temporal vector component $P$ exceeding the value $\sqrt{2}\Mpl$. 
Then, the stability condition $|P|<\sqrt{2}\Mpl$ is violated, as it happens in case (b).
Hence the hairy BHs that satisfy the stability requirements (\ref{sta1}) and (\ref{sta2}) on the horizon are generally prone to angular Laplacian instability at large distances.

The next-to-leading order contribution to $c_{\Omega,1}^2 (r=r_h)$ in Eq.~(\ref{cOho}) has 
the dependence proportional to $b_3 (r-r_h)$. 
This means that, for $b_3 \gg 1$, the distance $r_c$ at which $c_{\Omega,1}^2=0$ should be close to the horizon.
Incorporating the next-to-leading order term of $c_{\Omega,1}^2$ around $r=r_h$ and taking 
the limit $b_3 \gg 1$, we obtain 
\be
r_c \simeq r_h+\frac{\sqrt{2}f_1^{5/2}}
{b_3 \sqrt{f_1(a_1^2+2f_1)}[a_0 (2a_0+a_1)-f_1](2a_0+a_1)}r_h\,.
\ee
In the limit $b_3 \gg 1$, the critical distance $r_c$ approaches $r_h$. For larger $b_3$, the transition to the regime of negative instability ($c_{\Omega,1}^2<0$) occurs at the distance closer to the horizon.

In principle, in the absence of mathematical proof that $c_\Omega^2$ is bound to be negative in some regions of spacetime, we should be looking for the possible existence of parameter spaces in which the BH solutions are stable. 
In doing so, we have performed, for some fixed positive values of $\beta_3$, while keeping $f_1$ to be unity, a search for stable solutions in broad parameter space for the variables $(a_0,a_1)$, i.e.,\ $10^{-2}\leq a_0\leq10$, and $10^{-2}\leq | a_1|\leq10$. Note that this gives the same values of $c_{\Omega,1}^2$ as those derived by the rescaling of $f_{\infty}$ to 1. 
What we found is that: 1) either there is no background BH solution at all, or 2) the background BH solution exists but it resembles either case (a) or case (b) discussed above.

The above discussion shows that it is difficult for hairy BHs to satisfy all the three stability conditions (\ref{conL}), (\ref{sta1}), and (\ref{sta2}) associated with the absence of ghosts and angular Laplacian instabilities. 
Moreover, we also require the condition (\ref{s12}) to ensure that the perturbation is not subject to 
Laplacian instabilities along the radial direction. 
Even if the inequality (\ref{s12}) is satisfied, one of the dynamical perturbations has two squared 
propagation speeds $c_r^2$ whose magnitudes are different from each other. Such pathological 
behavior, together with the angular instability discussed above, makes it unlikely to realize stable hairy BHs in theories with $m=0$ and $\beta_3 \neq 0$.

\section{Stability of BHs for $m \neq 0$ and $\beta_3 \neq 0$}
\label{sta2sec}

We proceed to study the linear stability of BHs in theories with $m \neq 0$ and $\beta_3 \neq 0$. 
As we discussed in Sec.~\ref{nohairBHsec}, the presence of a mass term besides cubic vector Galileons leads to the no-hair BH solution given by 
\be
f=h=1-\frac{2M}{r}\,,\qquad 
A_0=0\,,\qquad A_1=0\,.
\ee
In this case, the gauge choice (\ref{gauge3}), i.e., 
\be
G=0\,,\qquad h_1=0\,,\qquad h_0=0\,,
\label{gauge3d}
\ee
is convenient for obtaining the stability conditions in the small-scale limit. 
To expand the action (\ref{action}) up to the second order in perturbations, we take into account the perturbation $G$ besides $H_0$, $H_1$, $H_2$, $K$, $\delta A_0$, $\delta A_1$, and $\delta A_2$. 
After performing the $\theta$ and $\varphi$ integrations for the quadratic-order action, we vary it with respect to eight perturbed fields and finally set $G=0$.
Combining the perturbation equation for $K$ with that for $G$, we obtain the following simple relation 
\be
H_0=H_2\,,
\ee
so that there are six fields $H_1$, $H_2$, $K$, $\delta A_0$, $\delta A_1$, and $\delta A_2$ left in the analysis. 

For one of the dynamical perturbations, we introduce 
\be
\psi \equiv K+\frac{2h}{L+1-3h}(H_2-rK')\,,
\ee
which allows us to express $H_2$ with respect to $\psi$, $K$, and $K'$. 
Then, the perturbation equations of motion for $H_0$ and $H_1$ are written, respectively, as 
\ba
& &
\left[ (L-2)r+6M \right] \left[LK+2(2M-r)\psi' \right]
-\left( L^2 r + 2LM + 8M - 4r \right)\psi=0\,,\label{H0eq}\\
& & 2L (r-2M) H_1+2r \left[ (L-2)r+6M \right] 
\dot{\psi}-2L r^2 \dot{K}=0\,.
\label{Keq}
\ea
From Eq.~(\ref{H0eq}), we can express $K$ in terms of $\psi$ and $\psi'$, so that $H_2$ depends on $\psi$, $\psi'$, and $\psi''$.
Furthermore, we can solve Eq.~(\ref{Keq}) for $H_1$ to express $H_1$ with respect to 
$\dot{\psi}$ and $\dot{\psi}'$.
Substituting $K$, $H_1$, $H_2$ and their derivatives into the second-order perturbed action, the resulting action contains $\psi$, $\delta A_0$, $\delta A_1$, $\delta A_2$, and their derivatives. 
Then, we vary the second-order action with respect to $\psi$, $\delta A_0$, $\delta A_1$, and $\delta A_2$ to obtain their perturbation equations of motion. 
The behavior of perturbations in the small-scale limit is known by assuming the solutions to the four perturbed fields as 
\be
\psi=\tilde{\psi} e^{-i (\omega t-kr)}\,,\qquad 
\delta A_0=\tilde{\delta A}_0 e^{-i (\omega t-kr)}\,,\qquad 
\delta A_1=\tilde{\delta A}_1 e^{-i (\omega t-kr)}\,,\qquad 
\delta A_2=\tilde{\delta A}_2 e^{-i (\omega t-kr)}\,,
\ee
where $\tilde{\psi}$, $\tilde{\delta A}_0$, $\tilde{\delta A}_1$, and $\tilde{\delta A}_2$ are 
assumed to be constant. 
The perturbation equation for $\delta A_0$ can be expressed as 
\be
\tilde{\delta A}_0=-\frac{[(r-2M)(kr-2i) 
\tilde{\delta A}_1+i L \tilde{\delta A}_2] \omega}
{(k^2+m^2) r^2-2k r(k M+i)+4kM i+L}\,.
\ee
We will exploit this relation to eliminate $\tilde{\delta A}_0$ from the field equations for $\tilde{\delta A}_1$ and $\tilde{\delta A}_2$, after which we are left with the three algebraic equations containing $\tilde{\psi}$, $\tilde{\delta A}_1$, and $\tilde{\delta A}_2$.

The field equation for $\tilde{\psi}$ is decoupled from the other two, so that it is expressed in the form 
\be
U_{11} \tilde{\psi}=0\,,
\ee
where $U_{11}$ is a $r$-dependent coefficient. The no-ghost condition is determined by picking up terms proportional to $\omega^2$ in $U_{11}$, which translates to 
\be
K_1 \equiv \frac{(L-2)\Mpl^2 r^2}{2L}>0\,,
\ee
which is automatically satisfied for $l \geq 2$. 
The propagation along the angular direction is known by taking the limit $\omega r_h \approx l \gg k r_h \gg 1$ in $U_{11}$. This gives the dispersion relation 
\be
r^3 \omega^2-L \left( r-2M \right)=0\,.
\label{disp1}
\ee
Since the squared angular propagation speed $c_{\Omega,1}^2$ is related to 
$\omega^2$ as $c_{\Omega,1}^2=r^2 \omega^2/(fl^2)=r^3 \omega^2/[l^2(r-2M)]$, Eq.~(\ref{disp1}) gives
\be
c_{\Omega,1}^2=1\,.
\ee
For the radial propagation, taking the limit $\omega r_h \approx k r_h \gg l \gg  1$ in $U_{11}$ leads to the dispersion relation 
\be
r^2 \omega^2-k^2 \left( r-2M \right)^2=0\,.
\label{disp2}
\ee
The squared radial propagation speed, which is defined by $c_{r,1}^2=(fh)^{-1}\omega^2/k^2
=r^2 \omega^2/[k^2(r-2M)^2]$, yields 
\be
c_{r,1}^2=1\,.
\ee
Thus, the perturbation $\psi$ has luminal propagation speeds in both angular and radial directions.

The other two modes $\tilde{\delta A}_1$ and $\tilde{\delta A}_2$ are coupled to each other. 
They are expressed in the following forms
\be
U_{22}\tilde{\delta A}_1+U_{23}\tilde{\delta A}_2=0\,,\qquad 
U_{32}\tilde{\delta A}_1+U_{33}\tilde{\delta A}_2=0\,.
\ee
The $r$-dependent coefficients $U_{22}$, $U_{23}$, $U_{32}$, and $U_{33}$ contain terms proportional 
to $\omega^2$. Expressing these coefficients in the forms $U_{ij}=U_{ij}^{\rm kin}\omega^2$, where $i$ and $j$ are either 2 or 3, we have 
\ba
U_{22}^{\rm kin} &=& \frac{(L+m^2 r^2) r^2}
{(k^2 r-2i k)(r-2M)+L+m^2 r^2}\,,\label{NG2no}\\
U_{22}^{\rm kin}U_{33}^{\rm kin}
-U_{23}^{\rm kin}U_{32}^{\rm kin} &=& 
\frac{L m^2 r^5}
{(r-2M)[(k^2 r-2i k)(r-2M)+L+m^2 r^2]}\,.\label{NG3no}
\ea
Taking the large $k$ limit in Eqs.~(\ref{NG2no}) and (\ref{NG3no}), the no-ghost conditions for the 
fields $\delta A_1$ and $\delta A_2$ are given by 
\ba
K_2 &\equiv& \frac{(L+m^2 r^2) r^2}
{k^2r(r-2M)+L+m^2 r^2}>0\,,\\
K_3 &\equiv& \frac{L m^2 r^5}
{(r-2M)[k^2r(r-2M)+L+m^2 r^2]}>0\,,
\ea
which are both satisfied outside the horizon ($r>2M$).

To derive the angular propagation speeds $c_{\Omega}$ of $\delta A_1$ and $\delta A_2$, we take the limit $\omega r_h \approx l \gg k r_h \gg 1$ in the determinant equation $U_{22}U_{33}-U_{23}U_{32}=0$. 
Replacing $\omega^2$ with $l^2 (r-2M) c_\Omega^2/r^3$ further, we obtain 
\be
\frac{r-2M}{r}L^2 m^2 \left( c_{\Omega}^2-1 
\right)^2=0\,,
\ee
which gives the following two solutions for $c_{\Omega}^2$, 
\be
c_{\Omega,2}^2=1\,,\qquad 
c_{\Omega,3}^2=1\,.
\ee
The radial propagation speeds $c_{r}$ of $\delta A_1$ and $\delta A_2$ are known by taking the limit 
$\omega r_h \approx k r_h \gg l \gg  1$ in the determinant equation $U_{22}U_{33}-U_{23}U_{32}=0$ 
and by replacing $\omega^2$ with $k^2 (r-2M)^2 c_r^2/r^2$. 
This leads to 
\be
(r-2M)^2 k^2 L m^2 
\left( c_r^2-1 \right)^2=0\,,
\ee
whose solutions to $c_r^2$ are given by 
\be
c_{r,2}^2=1\,,\qquad 
c_{r,3}^2=1\,.
\ee

From the above discussion, the three dynamical perturbations $\psi$, $\delta A_1$, and $\delta A_2$ have luminal propagation speeds in both angular and radial directions, without the appearance of ghosts.\footnote{We can arrive at the same conclusions by choosing a different gauge. 
For instance, in the gauge $L G=2K$, $h_0=0=h_1$, we can integrate out $K$ by using the perturbation equations of motion for $H_0$ and $H_1$. At this point, the field $H_2$ decouples from the other fields, $\delta A_0$, $\delta A_1$, and $\delta A_2$. Using the WKB approximation in the 
small-scale limit, we can integrate out $\delta A_0$ leaving only $\delta A_1$ and $\delta A_2$. 
Then, we find the luminal propagation for the three dynamical fields $H_2$, $\delta A_1$, and $\delta A_2$.} The cubic coupling $\beta_3$ does not appear anywhere in the perturbation equations of 
motion, by reflecting the fact that the background value of $X$ vanishes due to the no-hair property of BHs ($A_0=0=A_1$). 
This means that the stability of BHs in theories with $m \neq 0$ and $\beta_3 \neq 0$ is the same as that of the Schwarzschild BH (\ref{Schwarz}) in Einstein-Proca theories with $m \neq 0$ and $\beta_3=0$. 
As we showed in Sec.~\ref{sta1sec}, the hairy BHs arising in theories with $m=0$ and $\beta_3 \neq 0$ are generally prone to Laplacian instability. 
The result in this section shows that introducing the mass term $m$ to the cubic vector Galileon regularizes the solutions in such a way that the BHs are linearly stable against all of the three dynamical perturbations.
Thus, the mass term acts as a regulator in the behavior of perturbations as they become stable and standard. On top of that, the background of this massive cubic vector Galileon theory is the same as that in GR without the vector field.
This also implies that theories with $m \neq 0$ and $\beta_3 \neq 0$ do not require some screening mechanism of fifth forces on the SSS configuration. 

\section{Conclusions}
\label{consec}

We studied the linear stability of BHs on the SSS background for a subclass of GP theories given by the action (\ref{action}). The difference from Einstein-Maxwell theories is that the vector-field mass term $m$ and the cubic vector Galileon with a coupling constant $\beta_3$ break the $U(1)$ gauge symmetry.
For $m=0$, it is known that the cubic vector Galileon induces hairy BH solutions with nonvanishing temporal and longitudinal vector components $A_0$ and $A_1$. As we observe in Eq.~(\ref{back4d}), $A_1$ is divergent for $\beta_3 \to 0$, and hence the hairy BH is present only if $\beta_3 \neq 0$. In other words, in theories with $m=0$ and $\beta_3 \neq 0$, there is no continuous limit to the RN BH solution as $\beta_3 \to 0$.

In GP theories, there are in general three propagating DOFs in the even-parity sector.
The odd-parity sector contains two dynamical DOFs, whose behavior in the small-scale limit was already discussed in Ref.~\cite{Kase:2018voo}. 
For the odd-parity modes, the hairy BHs realized by cubic vector Galileons suffer from neither ghost nor Laplacian instabilities.
In this paper, we derived conditions for the absence of ghosts and Laplacian instabilities in the even-parity sector under the eikonal/WKB approximation. 
We showed that one of the even-parity dynamical perturbations has a nontrivial squared angular propagation speed $c_{\Omega,1}^2$ different from 1, while the other two dynamical modes have luminal propagation speeds. 
We also found that the radial propagation speed $c_r$ for one of the dynamical even-parity modes has two different values of $c_r^2$, whereas the other two are luminal.

For hairy BHs realized in theories with $m=0$ and $\beta_3 \neq 0$, we need to satisfy the two stability conditions (\ref{sta1}) and (\ref{sta2}) on the horizon together with the other inequality (\ref{conL}) at spatial infinity. 
For all the asymptotically flat hairy solutions we found, two different behaviors occur depending on 
the model parameters and boundary conditions. 
In one case ($\beta_3 a_0$ close to 0), an angular Laplacian instability on the horizon is associated with negative values of $c_{\Omega,1}^2$.
This horizon instability can be avoided for large $\beta_3 a_0$ satisfying the inequality (\ref{sta1}), but in such cases, $c_{\Omega,1}^2$ generally approaches negative values far away from the horizon. 
Indeed, we have not found any viable parameter space in which $c_{\Omega,1}^2$ is positive throughout the horizon exterior.
Taking into account the pathological behavior of $c_r^2$ mentioned above, the hairy BHs in theories with $m=0$ and $\beta_3 \neq 0$ are excluded as stable and healthy solutions. 

If we introduce the mass term besides the cubic vector Galileon ($m \neq 0$ and $\beta_3\neq 0$), the resulting background BH solution is the Schwarzschild geometry with vanishing temporal and longitudinal vector components.
In such cases, we showed that the no-ghost conditions are satisfied outside the horizon, with luminal speeds of propagation for three dynamical perturbations in the even-parity sector. Thus, the mass term $m$ regulates the unstable property of hairy BHs for cubic vector Galileons at the cost of removing the vector hair. Since the coupling $\beta_3$ does not appear in the perturbation equations of motion, the no-hair Schwarzschild BH for $m \neq 0$ and $\beta_3 \neq 0$ cannot be distinguished from that for $m \neq 0$ and $\beta_3=0$. 

In this paper, we focused on cubic vector Galileon theories with the coupling function $G_3(X)=\beta_3 X$.
If we consider more general cubic-order couplings $G_3(X)=\beta_3 X^n$ ($n \geq 2$) with $m=0$, it is known that there is another branch satisfying the relation $X=0$ \cite{Kase:2018voo}. 
Again, we cannot take the limit $\beta_3 \to 0$ for the existence of this branch. In such cases, the BHs may be subject to linear instabilities, as we found for cubic vector Galileons, but this deserves a separate study. 

If we consider power-law quartic couplings $G_4(X)=\beta_4 X^n$ with $n \geq 1$ in GP theories, the hairy BHs with $A_1 \neq 0$ are subject to instabilities in the odd-parity sector around the horizon \cite{Kase:2018voo}.
For power-law quintic couplings $G_5(X)=\beta_5 X^n$ with $n \geq 1$, the background BH solutions with $A_1 \neq 0$ have discontinuities outside the horizon \cite{Heisenberg:2017hwb}.
However, the $\ln |X|$ dependence in $G_5(X)$, which corresponds to vector-Gauss-Bonnet theories, allows the existence of hairy BHs even for $m \neq 0$ \cite{Aoki:2023jvt}. 
The sixth-order power-law couplings $G_6(X)=\beta_6 X^n$, which characterize the interaction between the vector field and the double dual Riemann tensor, give rise to hairy BHs with $A_1=0$ \cite{Heisenberg:2017xda, Heisenberg:2017hwb} that can be linearly stable against odd-parity perturbations \cite{Kase:2018voo}. For the specific $U(1)$-invariant coupling $G_6(X)=\beta_6={\rm constant}$ \cite{Horndeski:1976gi}, there are the range of $\beta_6$ in which all the linear 
stability conditions for electrically or magnetically charged BHs are satisfied \cite{Chen:2024hkm}.
It will be of interest to study the linear stability of BHs in full GP theories and classify the surviving models without theoretical pathologies.

\section*{Acknowledgements}

The work of ADF was supported by the JSPS Grants-in-Aid for Scientific Research No.~20K03969. 
RK is supported by the Grant-in-Aid for Scientific Research (C) of the JSPS No.~23K03421.
ST was supported by the Grant-in-Aid for Scientific Research Fund of the JSPS No.~22K03642 and Waseda University Special Research Project No.~2024C-474. 

\appendix

\section{Coefficients associated with $c_r$}

The quantities $s_1$ and $s_2$ in Eq.~\eqref{creq} are given by
\ba
s_1
&=&
\frac{2r\sqrt{h}\beta_3}{\sqrt{f}\,s_0}
\Bigg[
A_1 \bigg\{\Big(r \beta_3 A_1^{3} ( 12 A_1-A_1' r ) h^{2}-2 \Mpl^{2} (2 A_1' r +7 A_1 ) h -2 \Mpl^{2} A_1 (h' r -2)\Big) A_0 +r A_0' h A_1 (r \beta_3 h A_1^{3}-2 \Mpl^{2})\bigg\} \notag\\
&&
\times(r \beta_3 h A_1^{3}-2 \Mpl^{2}) h f^{5}
-A_0 \bigg\{\Big(r^{2} \beta_3^{2} ( 28 A_1-3 A_1' r ) A_1^{5} h^{3}-28 r \Mpl^{2} \beta_3 h^{2} A_1^{3}-2 (r^{2} A_1^{3} h' \beta_3 -4 A_1 \beta_3 ( \Mpl^{2}+A_1^{2}) r\notag\\
&&
-6 \Mpl^{2}) \Mpl^{2} h -8 \Mpl^{4}\Big) A_0^{2}-r A_0' (3 A_1^{6} r^{2} \beta_3^{2} h^{2}-4 r \Mpl^{2} \beta_3 h A_1^{3}+4 \Mpl^{4}) h A_0 +2 r h^{2} A_1^{2} (r \beta_3 h A_1^{3}-2 \Mpl^{2}) \notag\\
&&
\times(2 f' \Mpl^{2}+r A_0'^{2}-r A_1^{3} f' \beta_3 h)\bigg\} f^{4}
+r A_0^{3} \bigg\{\beta_3 \Big(r \beta_3 A_1^{3} ( 20 A_1-3 A_1' r ) h^{2}+2 \Mpl^{2} (A_1' r +5 A_1 ) h +4 \Mpl^{2} A_1 \Big) A_0^{2}\notag\\
&&
-r A_0' \beta_3 (11 r \beta_3 h A_1^{3}-8 \Mpl^{2}) A_1 h A_0 +\Big(2 r \beta_3 A_1^{3} (7 f' \Mpl^{2}+2 r A_0'^{2}) h -7 A_1^{6} r^{2} \beta_3^{2} f' h^{2}-4 \Mpl^{2} A_0'^{2} r -12 \Mpl^{4} f' \Big) h \bigg\} f^{3}\notag\\
&&
+r^{2} A_0^{5} \Big\{9 r A_1^{3} f' \beta_3 h +\beta_3 (A_1' r -4 A_1 ) A_0^{2}+9 r A_0' \beta_3 A_1 A_0 -2 r A_0'^{2}-6 f' \Mpl^{2}\Big\} \beta_3 A_1 h f^{2}\notag\\
&&
-A_0^{7} \beta_3^{2} r^{3} (5 A_1^{2} f' f h -A_0^{2} f' +2 A_0 A_0' f )\Bigg]\,,\\
s_2
&=&
\frac{1}{s_0}\Bigg[
2 \Big(\Mpl^{2} (h -1)-2 \beta_3 h^{2}A_1^{3} r \Big) \left(r \beta_3 hA_1^{3}-2 \Mpl^{2}\right)^{2} h f^{5}
-r f^{4} h \bigg\{2 \Big(r hA_1^{3}\Mpl^{2} \beta_3 (15 h -1)+4 \Mpl^{4} (3 h +1)\notag\\
&&
-2 r^{2} \beta_3^{2}A_1^{6} h^{3}\Big) \beta_3A_1A_0^{2}+\Big(\left(A_1 (hA_1^{2} f' +2A_0A_0' ) \beta_3 -A_0'^{2}\right) r -2 f' \Mpl^{2}\Big) \left(r \beta_3 hA_1^{3}-2 \Mpl^{2}\right)^{2} h \bigg\}\notag\\
&&
+hA_0^{2} f^{3} \beta_3 r^{2} \bigg\{2 h^{2} rA_1^{5} (A_1^{2} f' h r +A_0A_0' r +2A_0^{2}) \beta_3^{2}
+\Big(2 \Mpl^{2} \left((3 h +1)A_0^{2}A_1^{2}-4A_0^{2} \Mpl^{2}-4A_1^{2} h rA_0A_0' \right)\notag\\
&&
-rA_1^{4} h^{2} (10 f' \Mpl^{2}+rA_0'^{2})\Big) \beta_3 +4 \Mpl^{2}A_1 h (f' \Mpl^{2}+rA_0'^{2})\bigg\}
+f^{2} \beta_3^{2} r^{2}A_0^{4} \bigg\{2A_0^{2} \left((3 h -1) \Mpl^{2}-2 \beta_3 h^{2}A_1^{3} r \right)\notag\\
&&
+rA_1^{2} h^{2} (2 rA_0' \beta_3A_1A_0 +2 f' \Mpl^{2}-rA_0'^{2})\bigg\}
-f h \beta_3^{2} r^{3}A_0^{6} \Big\{(2A_1 (hA_1^{2} f' +A_0A_0' ) \beta_3 -A_0'^{2}) r -2 f' \Mpl^{2}\Big\}
\notag\\
&&+\beta_3^{3} h r^{4}A_0^{8}A_1 f'
\Bigg]\,,
\ea
where
\ba
s_0
&=&
f^{5}h \Big[2 \Mpl^{2} (h' r +h -1)-A_1^{2} h \beta_3 (A_1 h' +2 A_1' h ) r^{2}\Big] (r \beta_3 h A_1^{3}-2 \Mpl^{2})^{2} +\beta_3 h^{2} r^{3} f^{4} A_1^{3}\Big[r \beta_3 h A_1^{3} \big\{(6 A_1' A_0^{2} \beta_3
\notag\\
&&
+4 \beta_3 A_0' A_0 A_1 +A_0'^{2}) h +2 \beta_3 A_0^{2} A_1 h' \big\}-4 h^{2} A_0^{2} A_1^{4} \beta_3^{2}
-2 \Mpl^{2} (7 A_0^{2} A_1 h' +12 A_0^{2} A_1' h +6 A_0 A_1 A_0' h ) \beta_3 -4 A_0'^{2} h \Mpl^{2}\Big]
\notag\\
&&
-\Mpl^{2} h r f^{4} \Big[2 \big\{13 h^{2} A_0^{2} A_1^{4} \beta_3^{2}+2 \Mpl^{2} h \big(2 A_0^{2} A_1^{2} \beta_3^{2}-2 (3 A_1' A_0^{2}+A_0' A_0 A_1 ) \beta_3 -A_0'^{2}\big)-h A_0^{2} A_1^{4} \beta_3^{2}-10 \beta_3 A_0^{2} \Mpl^{2} A_1 h' \big\} r
\notag\\
&&
-8 \beta_3 (7 h -1) A_0^{2} \Mpl^{2} A_1 \Big]
-h \Big[r^{2}\beta_3 h^{2} A_1^{3} \big\{2 (h A_1^{3} f' +3 A_1' A_0^{2}+8 A_0' A_0 A_1 ) \beta_3 +A_0'^{2}\big\}-2 \big\{\big(2 h^{2} A_1^{4} \beta_3
\notag\\
&&
+\Mpl^{2} (3 A_1 h' +8 A_1' h )\big) \beta_3 A_0^{2}+2 h \Mpl^{2} (h A_1^{3} f' \beta_3 +6 \beta_3 A_0' A_0 A_1 +A_0'^{2})\big\} r -2 \beta_3 A_0^{2} \Mpl^{2} (7 h +1) A_1 \Big] A_0^{2} \beta_3 r^{2} f^{3} A_1
\notag\\
&&
+f^{2} A_0^{4} r^{2} \beta_3^{2} \Big[2 A_0^{2}\big\{h A_1^{2} r \beta_3 ((-h' r +2 h ) A_1 +h r A_1' )+\Mpl^{2} (h' r +5 h -1)\big\} +4 r A_0' (5 r \beta_3 h A_1^{3}-3 \Mpl^{2}) h A_0 
\notag\\
&&
+r A_1^{2} h^{2} (6 r A_1^{3} f' \beta_3 h -12 f' \Mpl^{2}-r A_0'^{2})\Big] 
+r^{3} A_0^{6} \Big[A_1 \beta_3 r (-6 h^{2} A_1^{2} f' +A_0^{2} h' )-\big\{4 A_0 \beta_3 (2 r A_0' +A_0 ) A_1
\notag\\
&&
 -r A_0'^{2}-8 f' \Mpl^{2}\big\} h \Big] \beta_3^{2} f +2 \beta_3^{3} h r^{4} A_0^{8} A_1 f'
\,.
\ea
%

\bibliographystyle{mybibstyle}
\bibliography{bib}

\end{document}